\documentclass{elsart5p}
\usepackage{lineno}
\usepackage{amsfonts}
\usepackage{amssymb}
\usepackage{amsmath}
\usepackage{epsfig}
\usepackage{graphicx}
\usepackage{color}

\begin{document}

\begin{frontmatter}

\title
{Emergence and Stability of Vortex Clusters in Bose-Einstein Condensates:\\
a Bifurcation Approach near the Linear Limit}

\author[germany]{S.\ Middelkamp},
\author[pgk]{P. G.\ Kevrekidis},
\author[djf]{D. J.\ Frantzeskakis},
\author[sdsu]{R.\ Carretero-Gonz\'{a}lez\corauthref{cor1}}
\corauth[cor1]{Corresponding author}, and
\author[germany]{P.\ Schmelcher}

\address[germany]{%
Zentrum f\"ur Optische Quantentechnologien, Universit\"at
Hamburg, Luruper Chaussee 149, 22761 Hamburg, Germany
}
\address[pgk]{%
Department of Mathematics and Statistics, University of
Massachusetts, Amherst MA 01003-4515, USA
}
\address[djf]{%
Department of Physics, University of Athens, Panepistimiopolis,
Zografos, Athens 157 84, Greece
}
\address[sdsu]{%
Nonlinear Dynamical Systems Group{$^1$},
Computational Sciences Research Center, and\\
Department of Mathematics and Statistics,
San Diego State University, San Diego,
CA 92182-7720, USA
}
\thanks[nlds]{{\tt URL:} http://nlds.sdsu.edu/}
%


\begin{abstract}
We study the existence and stability properties
of clusters of alternating charge vortices
in Bose-Einstein condensates. It is illustrated
that such states emerge from cascades of symmetry-breaking
bifurcations that can be analytically tracked near the linear limit
of the system via weakly nonlinear few-mode expansions. We present the
resulting states that emerge near the first few eigenvalues of
the linear limit, and illustrate how the nature
of the bifurcations can be used to understand their stability. 
Rectilinear, polygonal and diagonal vortex
clusters are only some of the obtained states while mixed
states, consisting of dark solitons and vortex clusters, are
identified as well.
\end{abstract}

\maketitle



\end{frontmatter}

\section{Introduction}
\label{Sec1}

Vortices are among the most striking characteristics
of nonlinear field theories in higher-dimensional settings \cite{Pismen}.
They constitute one of the remarkable features of superfluids, while
playing also a key role in critical current densities and
resistances of type-II superconductors
through their transport properties, and are associated with quantum
turbulence in superfluid helium~\cite{donnelly}.
Vortices appear also in a wide variety of fields, ranging from fluid dynamics
\cite{chorin} to atomic physics \cite{emergent} and  optical physics \cite{optvort}.

A pristine setting for the study of vortices at the mesoscale
has emerged
after the realization of atomic Bose-Einstein condensates (BECs). In this context,
so-called {\it matter-wave vortices} were
experimentally observed therein~\cite{Matthews99}, by using a
phase-imprinting method between two hyperfine
spin states of a $^{87}$Rb BEC \cite{Williams99}. This achievement
subsequently
triggered extensive studies concerning
vortex formation, dynamics and interactions. For instance,
stirring the BECs \cite{Madison00} above a certain critical angular speed
\cite{Recati01,Sinha01,corro,Madison01} led to
the production of few vortices \cite{Madison01}, and even of
robust vortex lattices \cite{Raman}.
Vortices can also be formed in experiments by means of other
techniques,
such as by dragging obstacles through the BEC \cite{kett99} or by the nonlinear interference
of different condensate fragments \cite{BPAPRL}. Not only unit-charged, but also
higher-charged vortices were observed \cite{S2Ket} and their dynamical instabilities have been analyzed.

While a considerable volume of work has been dedicated to individual
vortices and to vortex lattices, arguably,
vortex clusters consisting of only a few vortices have attracted less interest.
The latter theme has become a focal point recently,
through the experiments involving two-vortex states (alias vortex dipoles) \cite{BPA_recent,dshall_recent},
as well as three-vortex states \cite{bagnato}. In Ref.~\cite{BPA_recent},
vortex dipoles were produced by dragging a localized light beam with appropriate speed through the
BEC, while in Ref.~\cite{dshall_recent} they
were distilled through the Kibble-Zurek mechanism \cite{kibblezurek}, previously proposed and
realized for vortices in Ref.~\cite{bpa_previous}. 
For the nonlinear
dynamics of the vortices in the dipoles of Ref.~\cite{dshall_recent},
see also the very recent analysis of Ref.~\cite{marchfinns}.
In Ref.~\cite{bagnato}, different types of three-vortex
configurations were produced by applying an external quadrupolar magnetic field on the BEC.
The principal ones among them were an aligned vortex ``tripole''
with a vortex of one topological charge straddling
two other oppositely charged vortices, and an equilateral triangle
of three same charge vortices.
On the theoretical side, few vortex states have been considered
also in a number of works. It was shown, in particular, that vortex
dipoles (consisting of a pair of vortices with opposite circulation) are
fairly robust in BECs
\cite{crasovan}. More elaborate
states, such as dipoles, tripoles and quadrupoles, were considered in
Refs.~\cite{mott1,mott2}. Dynamics of such few vortex states in the
weakly-interacting limit were performed in Ref.~\cite{klein}, while
the recent work of Ref.~\cite{komineas} connected the vortex dipoles
to the instability of dark soliton stripes; see also the important
earlier work of Ref.~\cite{pantoflas}.

In this work, we present a unifying analysis of
the existence and stability of
vortex clusters
(consisting of alternating charge vortices), by corroborating theoretical
investigations and numerical computations.
Our study
in Section \ref{Sec2}
will be based on the low-density limit
of near-linear excitations, where we will illustrate how they
emerge (bifurcate through symmetry-breaking bifurcations)
from states of the two-dimensional (2D) quantum harmonic oscillator.
Then, such theoretically identified states will
be continued via numerical computations in Section \ref{Sec3} to the strongly nonlinear
regime. From these continuations, we will be able to
infer numerous previously undiscovered vortex cluster states,
and to elucidate their stability properties (as well as compare
to the theoretical predictions). Lastly, in Section \ref{Sec4}, we will
summarize our findings and present some directions for future work.

\section{Model And Theoretical Analysis}
\label{Sec2}

We consider a quasi-2D (alias ``disk-shaped'') condensate confined in
a highly anisotropic trap with frequencies $\omega_z$ and $\omega_{\perp}$
along the transverse and in-plane directions, respectively. In the case
$\omega_{\perp} \ll \omega_z$ and $\mu \ll \hbar \omega_z$ (where $\mu$ is the chemical potential),
and for sufficiently low temperatures, the in-plane part $u(x,y,t)$
of the macroscopic BEC wave function obeys the following $(2+1)$-dimensional
Gross-Pitaevskii equation (GPE) (see, e.g., Ref.~\cite{emergent}):
\begin{equation}
i \hbar \partial_t u = \left[ - \frac{\hbar^{2}}{2m} \nabla_{\perp}^{2} + V(r)
+ g_{2D} |u|^{2} - \mu \right] u,
\label{veq1}
\end{equation}
where $\nabla_{\perp}^2$ is the in-plane Laplacian,
while the potential is given by $V(r)=(1/2)m \omega_{\perp}^{2}r^2$ (where $m$ is the atomic mass).
The effective 2D nonlinearity strength is given by $g_{2D}=g_{3D}/\sqrt{2\pi} a_{z} = 2\sqrt{2\pi} a a_{z} \hbar \omega_{z}$,
with $g_{3D} = {4 \pi \hbar^2 a}/{m}$, $a$ and $a_z = \sqrt{\hbar/m \omega_z}$ denoting, respectively,
the three-dimensional (3D) interaction strength, the $s$-wave scattering length, and the 
transverse harmonic oscillator length.
Equation (\ref{veq1}) can be expressed in the following dimensionless form,
\begin{equation}
i \partial_t u  = \left[ - \frac{1}{2} \nabla^2 + V(r)
+ |u|^{2} - \mu \right] u,
\label{veq1b}
\end{equation}
where the density $|u|^2$, length, time and energy are respectively measured
in units of $(2\sqrt{2\pi}a a_z)^{-1}$, $a_z$, $\omega_z^{-1}$ and
$\hbar\omega_z$. Finally, the harmonic potential is now given by $V(r)=(1/2) \Omega^2 r^2$,
with $\Omega = \omega_{\perp}/\omega_z$. From here on, all equations will be
presented in dimensionless units for simplicity.

Below, we will analyze the existence and linear stability of the nonlinear modes of
Eq.~(\ref{veq1b}). Notice that numerically the relevant nonlinear states will be identified as a
function of the chemical potential $\mu$ by means of a fixed point
(Newton iteration) scheme over a rectangular two-dimensional grid with suitably
small spacing. We will also explore the linear (spectral) stability of the obtained states
by means of the Bogoliubov-de Gennes (BdG) analysis.
The latter involves the derivation of the BdG equations, which stem from a linearization
of the GPE (\ref{veq1b}) around the stationary solution $u_0(x,y)$ via the ansatz
\begin{eqnarray}
u =u_0(x,y)
+ \left[a(x,y) e^{i \omega t}
+ b^{\ast}(x,y) e^{-i \omega^{\ast} t} \right],
\end{eqnarray}
where $\ast$ denotes complex conjugate. The solution of the ensuing BdG eigenvalue
problem yields the eigenfunctions $\{a(x,y),b(x,y)\}$ and eigenfrequencies $\omega$. Due
to the Hamiltonian nature of the system, if $\omega$ is an eigenfrequency
of the Bogoliubov spectrum, so are $-\omega$, $\omega^{\ast}$ and $-\omega^{\ast}$.
Notice that a linearly stable configuration is
tantamount to ${\rm Im}(\omega) =0$, i.e., all eigenfrequencies being real.
It is important to mention that in what follows we only resolve the relevant 
eigenvalues up to $10^{-2}$ due to computational domain constraints
and therefore all eigenvalues smaller than $10^{-2}$ will be omitted
in the figures.

Within the BdG analysis, a relevant quantity to consider is the {\em norm} 
$\times$ {\em energy} product
of a normal mode with eigenfrequency $\omega$, namely,
%
\begin{equation}
E=\int{dxdy(|a|^2-|b|^2)}
\omega.
\label{energy}
\end{equation}
The sign of this quantity, known as {\it Krein sign} \cite{MacKay},
is a topological property of each eigenmode.
In particular, if this sign is negative and
such a mode becomes resonant with a mode with positive Krein signature
then, typically, complex frequencies appear in the excitation
spectrum, i.e., a dynamical instability arises \cite{MacKay}.
We refer to this as
an oscillatory instability. Furthermore, dynamical instabilities
may arise due to a real mode eigenfrequency 
becoming imaginary. This typically coincides with 
a bifurcation of a new state.

\begin{figure*}[htb]
\centering
\begin{tabular}{cc}
\begin{tabular}{c}
\includegraphics[width=4.5cm,angle=0]{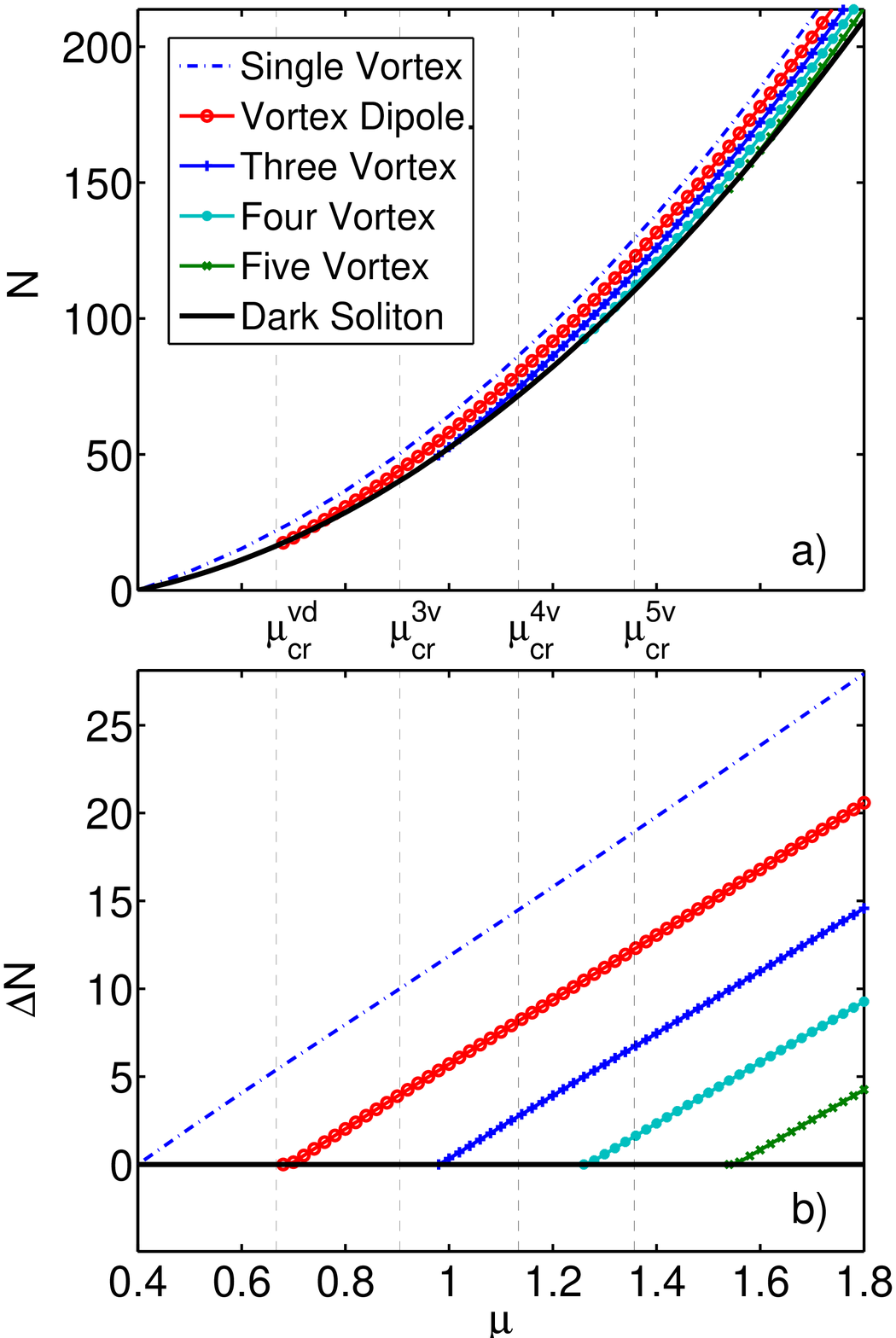}
\end{tabular}
&
\hskip0.5cm
\begin{tabular}{cc}
\includegraphics[width=3.5cm,angle=0]{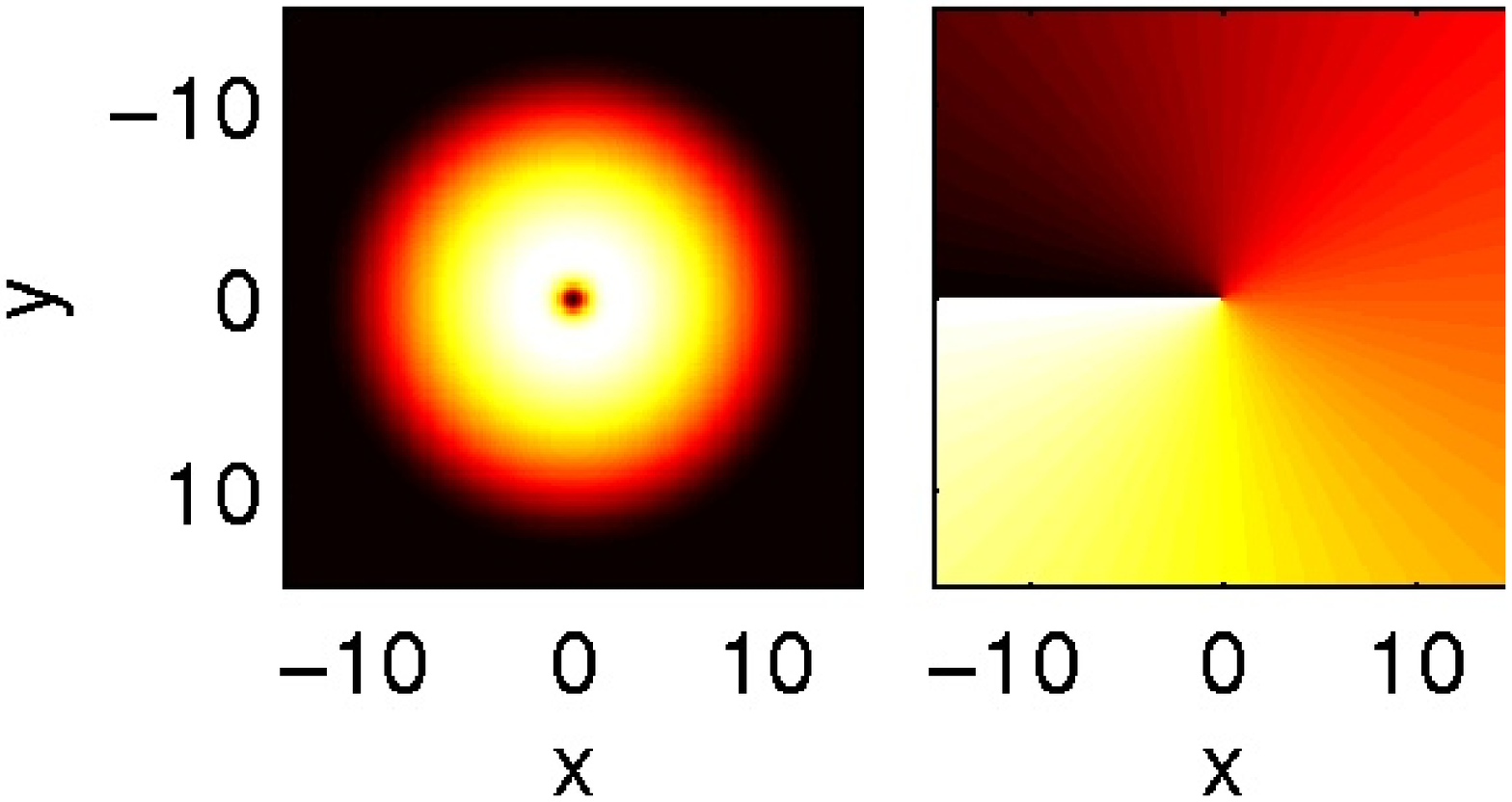} &
\includegraphics[width=3.5cm,angle=0]{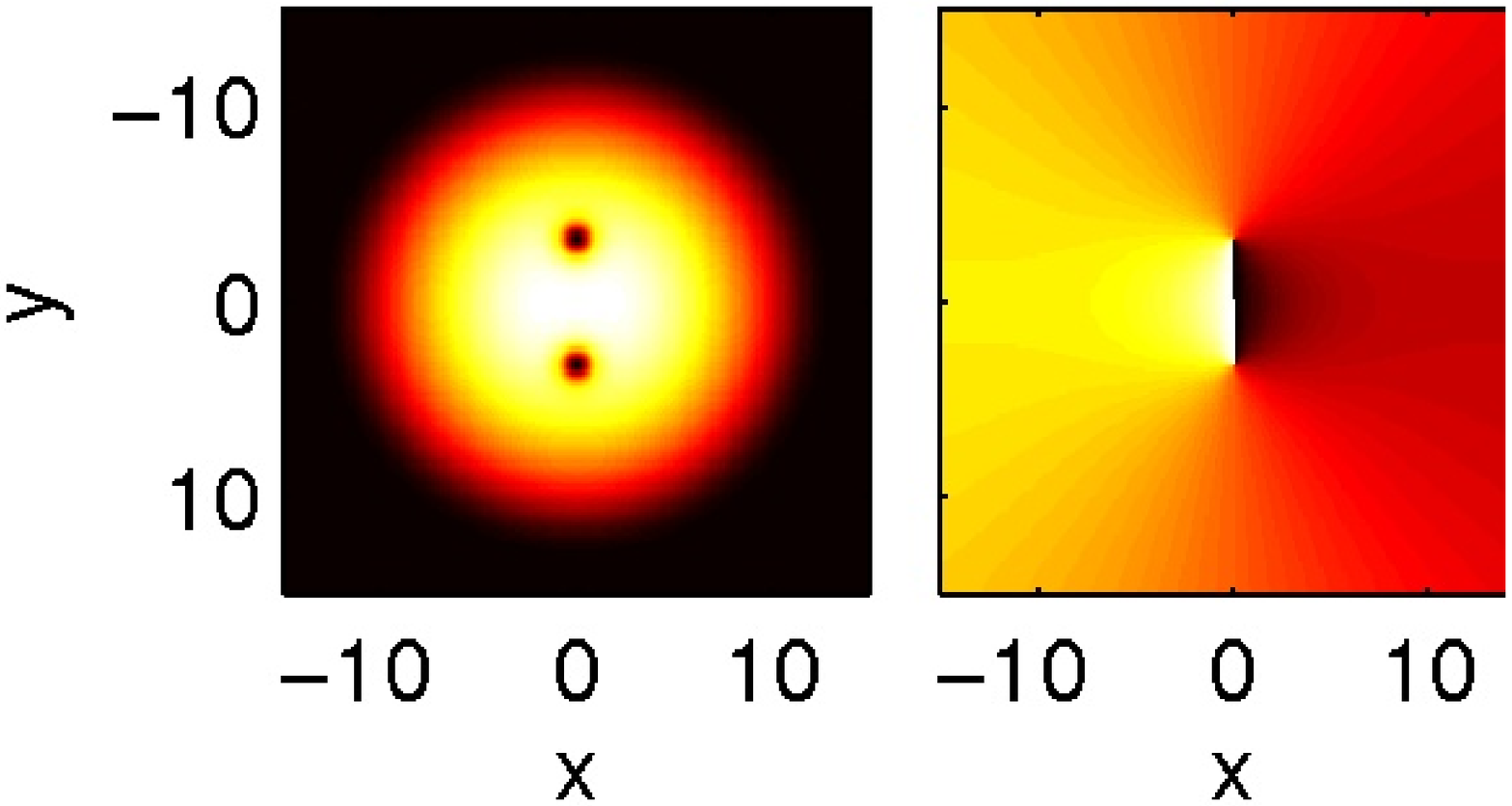}\\ 
\includegraphics[width=3.5cm,angle=0]{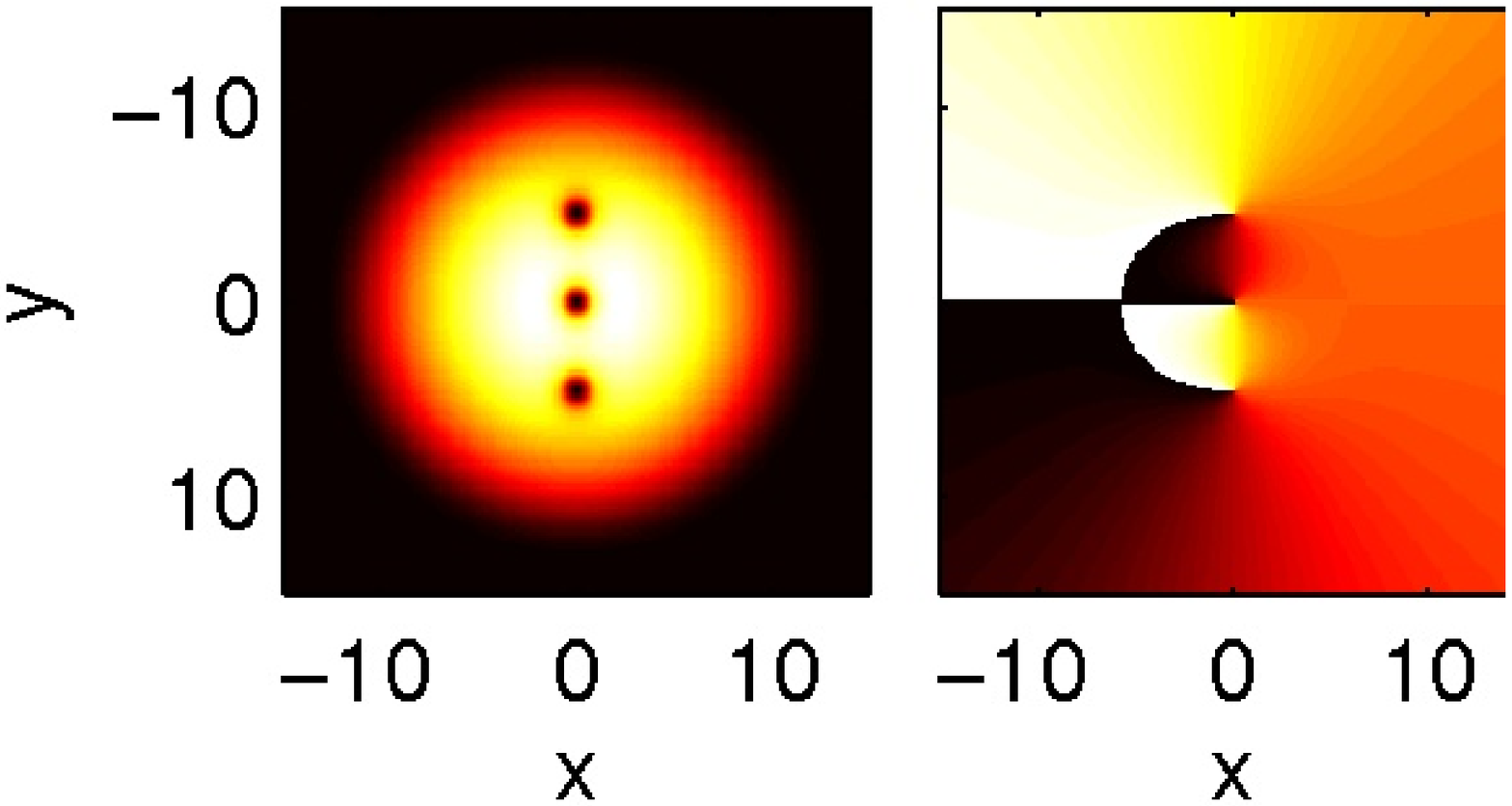}&
\includegraphics[width=3.5cm,angle=0]{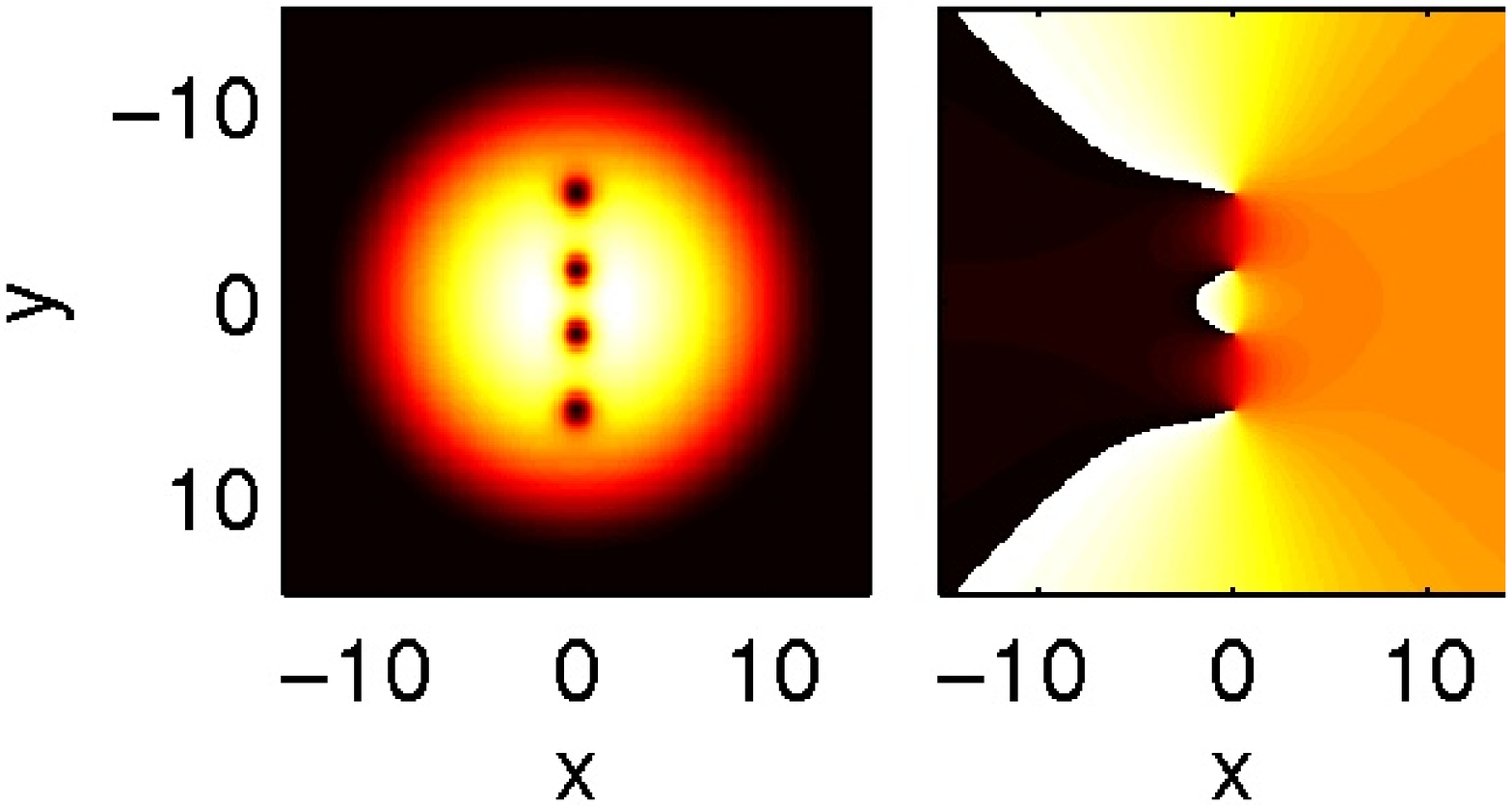}\\ 
\includegraphics[width=3.5cm,angle=0]{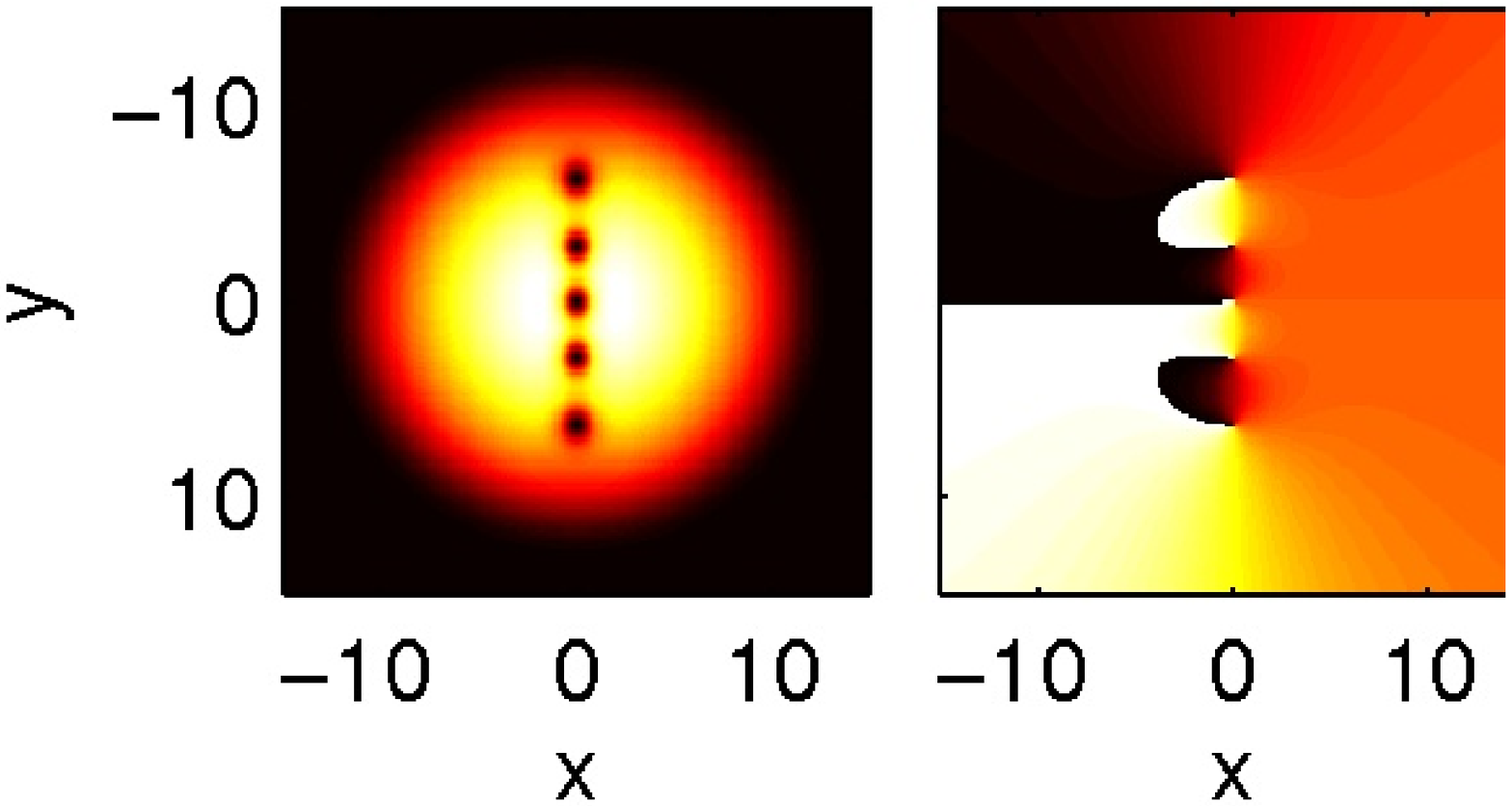}&
\includegraphics[width=3.5cm,angle=0]{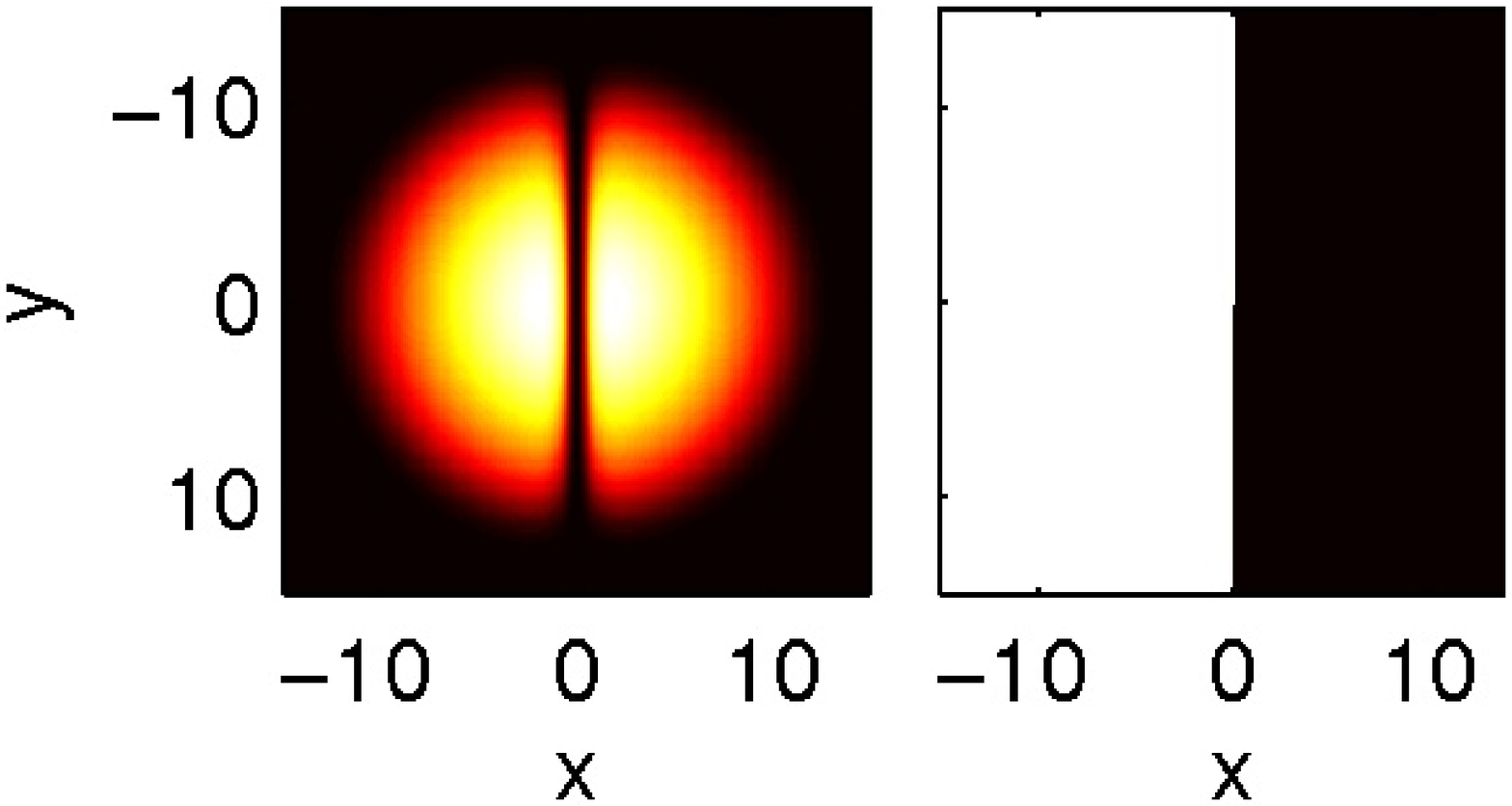}
\end{tabular}
\end{tabular}
\vspace{-0.2cm}
\caption{(Color online) 
Top left panel (a): Number of atoms as a function of the chemical
potential for the different states for $\Omega=0.2$ bifurcating
from the dark soliton stripe. 
Bottom left panel (b): Corresponding atom number difference with 
respect to the dark soliton stripe branch. 
Bifurcation of vortex multipole states arises when their atom
number difference from the dark soliton stripe
vanishes. The corresponding
theoretical predictions of such bifurcations are shown by the
vertical lines.
Right panels (from left to right and top to bottom): 
Density and phase profiles (left and right subpanels respectively)
corresponding to bifurcating states from 
the dark soliton stripe (bottom right panel):
the single vortex state, 
the aligned two- (vd), 
three- (3v), 
four- (4v) 
and 
five-vortex (5v) states.
}
\label{fig_DSS}
\end{figure*}

In the context of Eq.~(\ref{veq1b}), it is useful
to consider
the low-density (linear) limit. There, eigenstates of the 2D quantum harmonic oscillator
arise in the form $u_{nm}(x,y,t)=\exp(-i \mu t) H_n(x) H_m(y)$
(as well as linear combinations thereof), where $\mu=\Omega (n + m + 1)$
and $n$, $m$ quantify the order (and number of nodal lines) in each
direction. This produces a linear limit whose first excited state
has $\mu=2 \Omega$ and linear eigenstates $u_{10}$ and $u_{01}$. Notice
that one of their interesting linear combinations is $u_{10}+i u_{01}$,
which creates the single-charged vortex even at this linear limit;
this state exists for all higher values of $\mu$ and is shown in Fig.~\ref{fig_DSS},
but we will not be concerned with it further herein, as our focus
will be on clusters of vortices.

Each of the above mentioned linear eigenstates, $u_{10}$ and $u_{01}$, represents
a ``stripe'' i.e., a state with a nodal line. These can be
continued for higher chemical potentials $\mu$. As $\mu$ increases,
this state develops into
a one-dimensional (1D) dark soliton stripe, which is an exact 
analytical solution of Eq.~(\ref{veq1b})
in the absence of the trap~\cite{djf}. However, it is well-known that 
such a state is dynamically
unstable
towards decay into vortex structures \cite{kuzne,Feder:00-BPA:01,ander}. This decay can be understood
from a symmetry-breaking bifurcation point of view \cite{komineas,vd}.
In particular, it is possible to consider a two-mode (Galerkin-type) expansion,
similar to the one used in the literature of double-well potentials
(see e.g.~Ref.~\cite{ourDW}) in the form:
\begin{eqnarray}
u(x,y,t)= c_0(t) \phi_{0}(x,y) + c_1(t) \phi_{1}(x,y),
\label{eqn1}
\end{eqnarray}
%
where $c_{0}(t)$, $c_1(t)$ are complex time-dependent prefactors, while
$\phi_0=u_{10}(x,y)$, $\phi_1(x,y)=u_{0m}$ and $m>1$.
The resulting equations and analysis are formally equivalent to
the ones derived in Ref.~\cite{ourDW} (see Eqs.~(4)-(5) therein),
with appropriate modifications of the inner
products, but also with a {\it fundamental} difference. In the 1D double-well
setting, only symmetry-breaking bifurcations of asymmetric real 
solutions are predicted
(the so-called $\pi$-states that have recently been 
experimentally observed in Ref.~\cite{markusPRL}). The richer 2D case
enables bifurcations even when the relative phase $\Delta \phi$ between
the complex order parameters $c_0$ and $c_1$ is $\pi/2$.
In particular, such bifurcations are generically predicted at an atom number:
\begin{equation}
N_{\rm cr}=\frac{\omega_{0}-\omega_{1}}{I_{0}-I_1},
\label{Ncr}
\end{equation}
where $\omega_0$, $\omega_1$ are the linear state eigenvalues
corresponding to $\phi_0$ and $\phi_1$, while
$I_0=\int{\phi_{0}^2\phi_{1}^2dxdy}$ and $I_1=\int{\phi_{0}^4dxdy}$;
the critical chemical potential is
given by $\mu_{\rm cr}=\omega_{0} + I_1 N_{{\rm cr}}$.

\section{Numerical Results and Comparison with Theory}
\label{Sec3}

The above mentioned two-mode theory provides explicit predictions for the bifurcation
not only of the vortex dipole (vd) state when $m=2$, but also for an aligned
vortex tripole (3v) state (cf.~the experimental observations of Ref.~\cite{bagnato}) for
$m=3$, for an aligned vortex quadrupole (4v) state with $m=4$, etc. 
In fact, there is an entire cascade of such bifurcations, as $m$ increases, which 
occur progressively at 
$\mu_{\rm cr}^{\rm vd}=10 \Omega/3$, 
$\mu_{\rm cr}^{\rm 3v}=86 \Omega/19$, 
$\mu_{\rm cr}^{\rm 4v}=890 \Omega/157$, and 
$\mu_{\rm cr}^{\rm 5v}=726\Omega/107$ 
for $m=2,3,4,5$ etc., respectively. 
Notice that the number of vortices of the resulting 
cluster is evident by the number of 
intersections of the single nodal line of $u_{10}$ with the $m$
perpendicular nodal lines of $u_{0m}$, 
as well as the $\pi/2$ relative
phase of their complex prefactors
at these $m$ intersections. Also, it is evident that
that the sign changing of $u_{0m}$ at these intersections 
leads to an alternation of the ensuing vortex charges.
Importantly, general bifurcation
theory can be used to identify the stability characteristics of
the resulting states. In particular, since the stripe is dynamically
stable as it emerges from the linear limit, the vortex dipole
state that arises from it upon the first symmetry-breaking ``event''
($m=2$) should inherit this stability. However, now, once the
stripe has become unstable, all higher bifurcations with $m \geq 3$
will {\it necessarily} result into dynamically unstable states.

\begin{figure*}[htb]
\centering
\begin{tabular}{cc}
\begin{tabular}{c}
\includegraphics[width=4.5cm,angle=0]{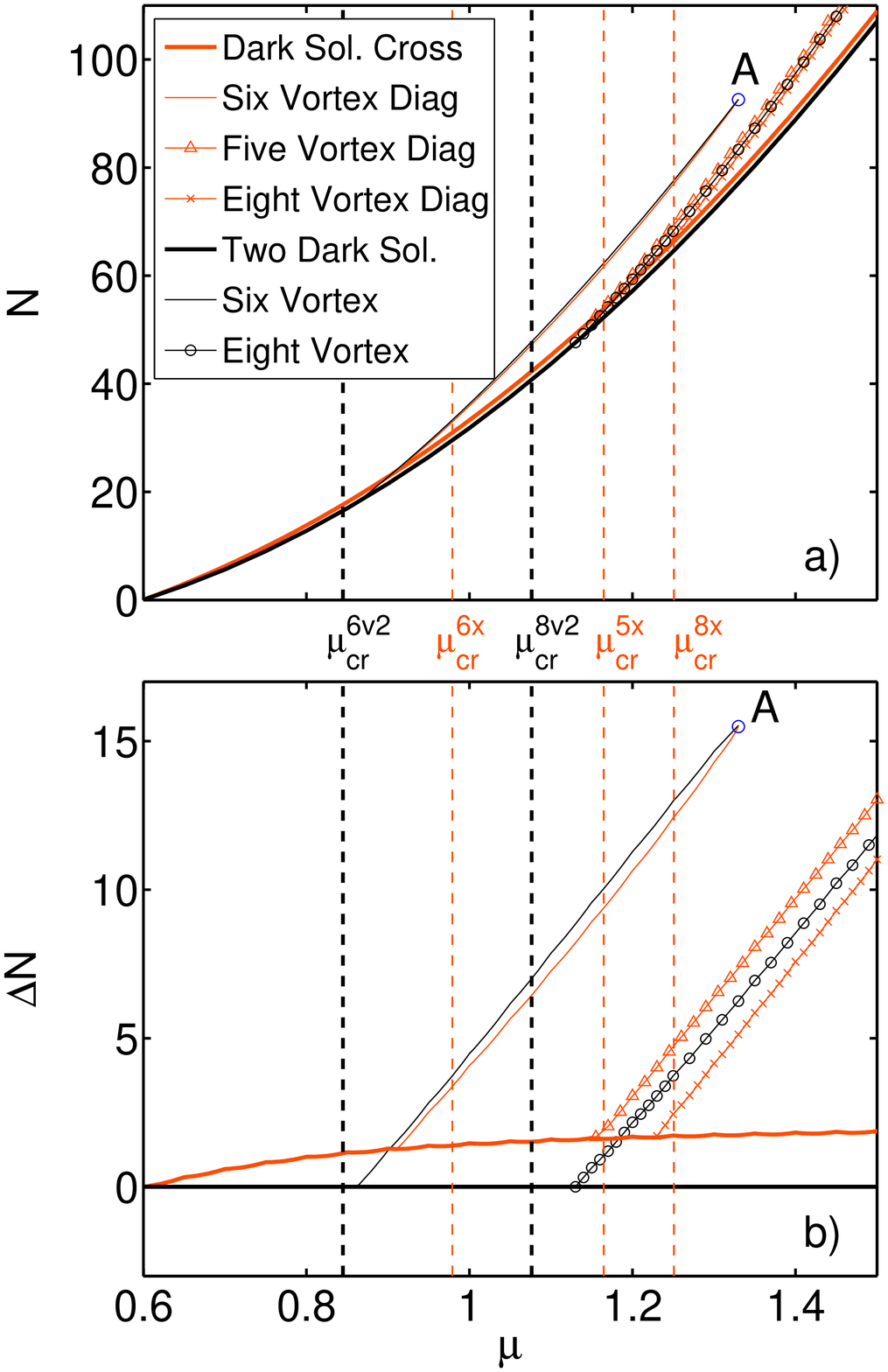}
\end{tabular}
&
\hskip0.6cm
\begin{tabular}{cc}
\includegraphics[width=3.5cm,angle=0]{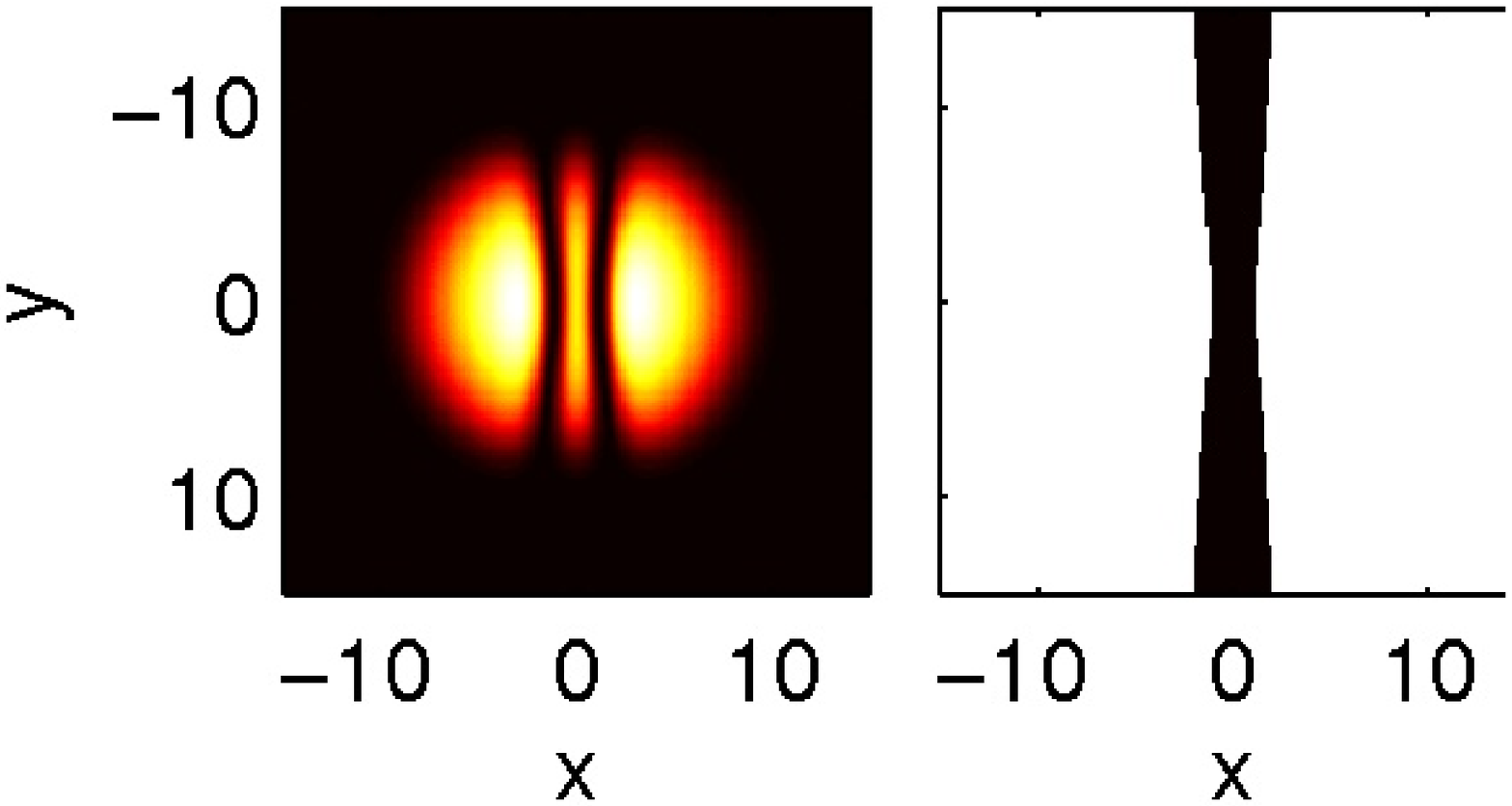} &
\includegraphics[width=3.5cm,angle=0]{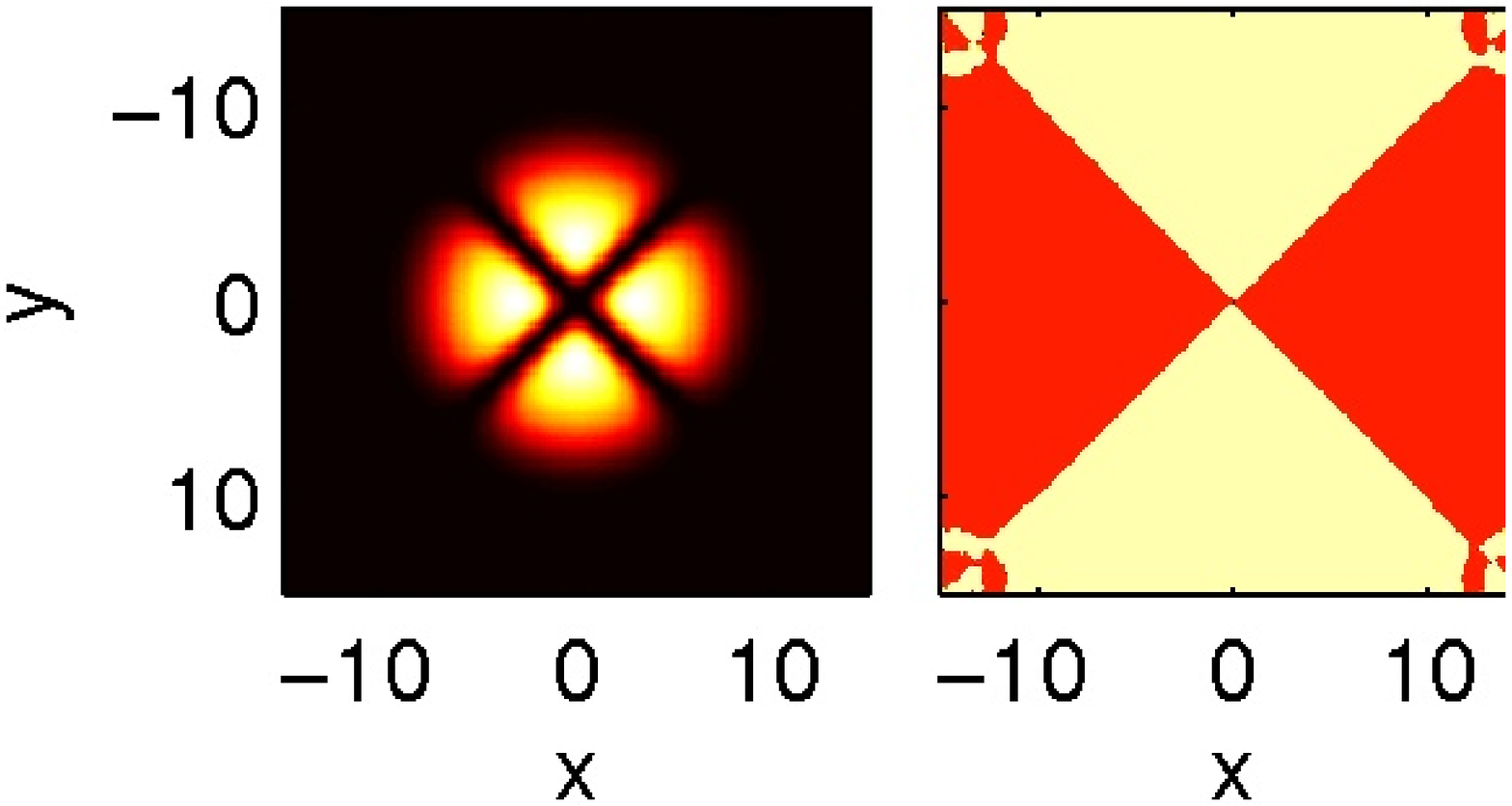} \\
\includegraphics[width=3.5cm,angle=0]{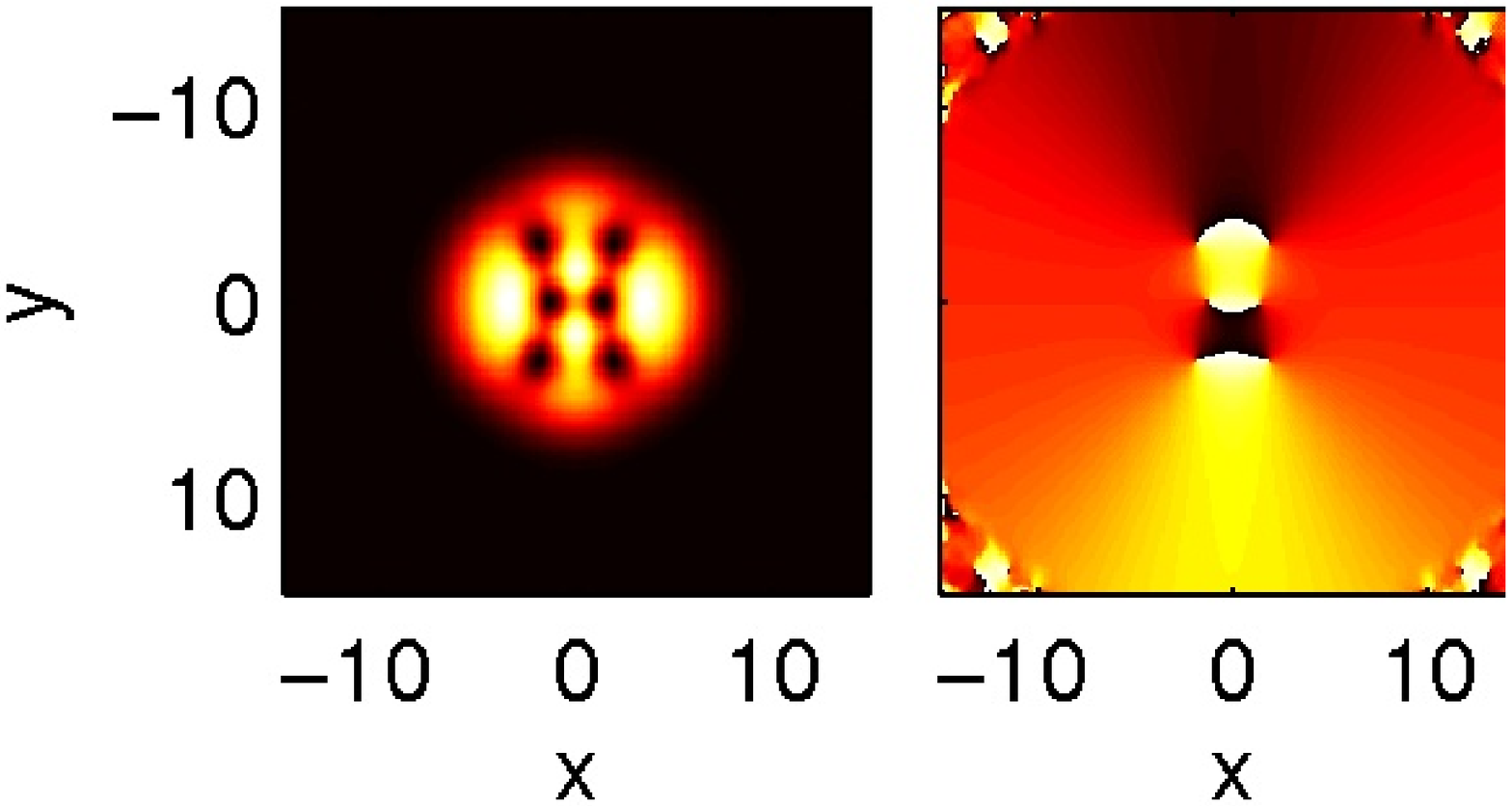} &
\includegraphics[width=3.5cm,angle=0]{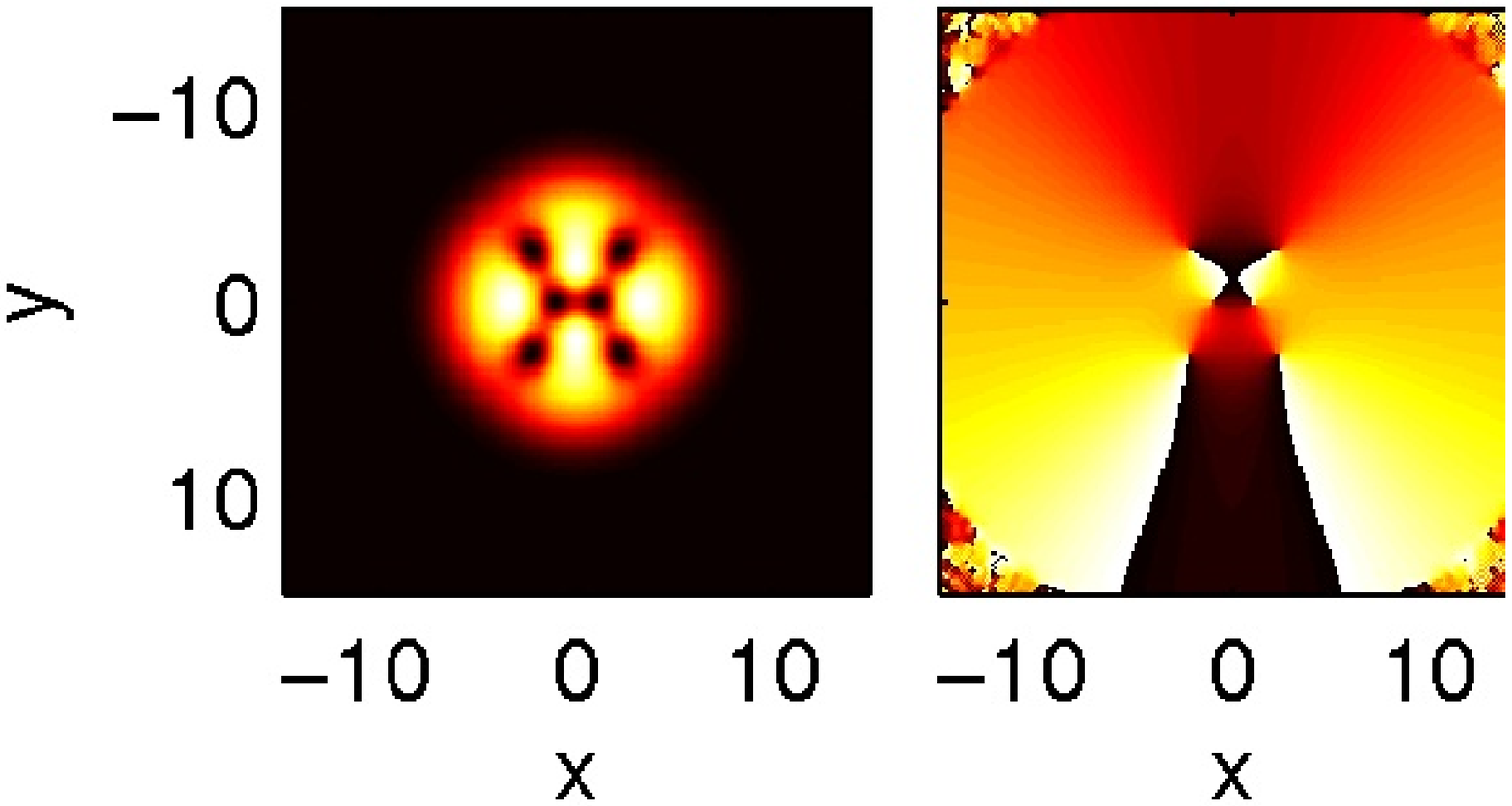} \\
\includegraphics[width=3.5cm,angle=0]{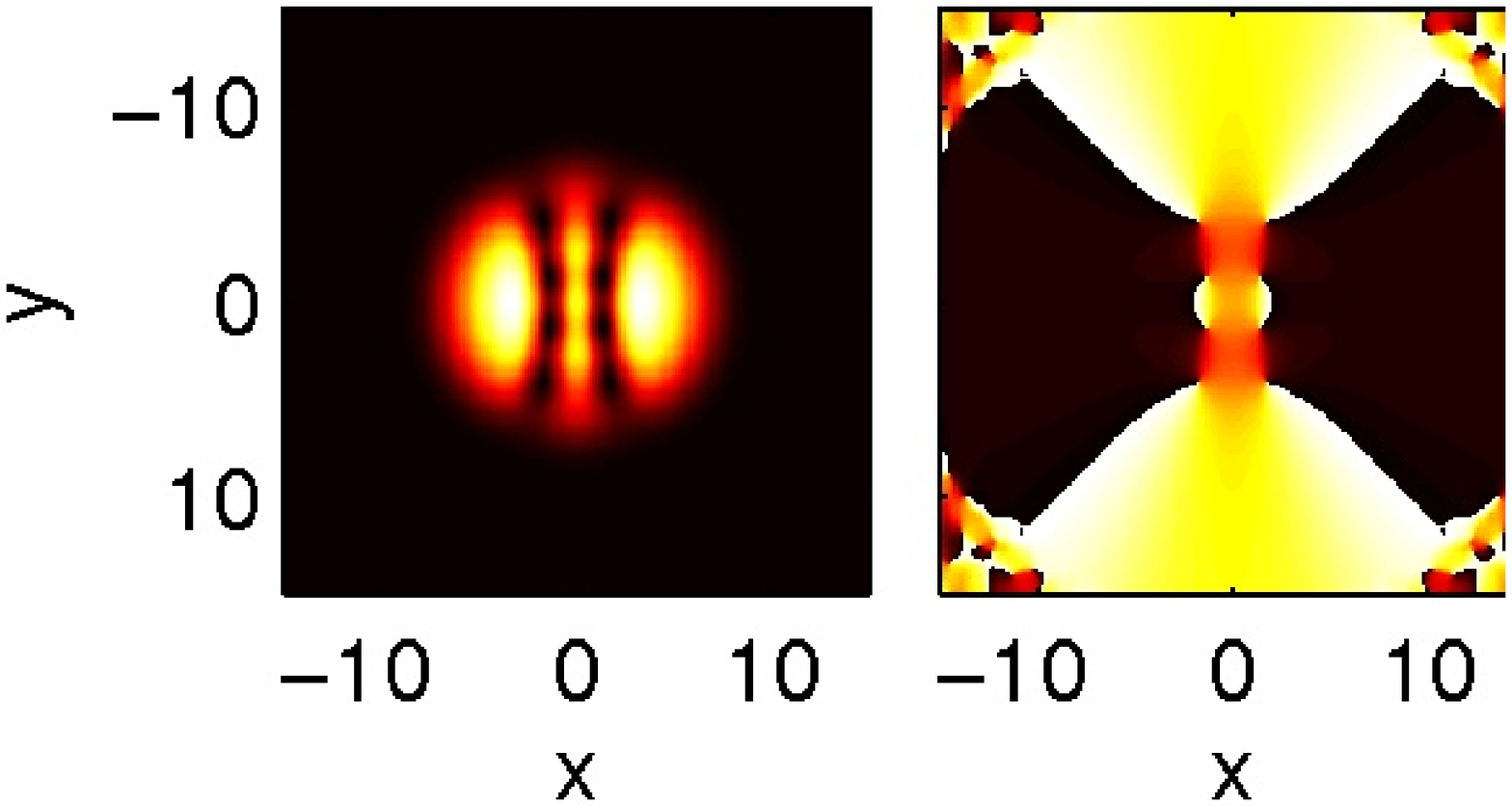} &
\includegraphics[width=3.5cm,angle=0]{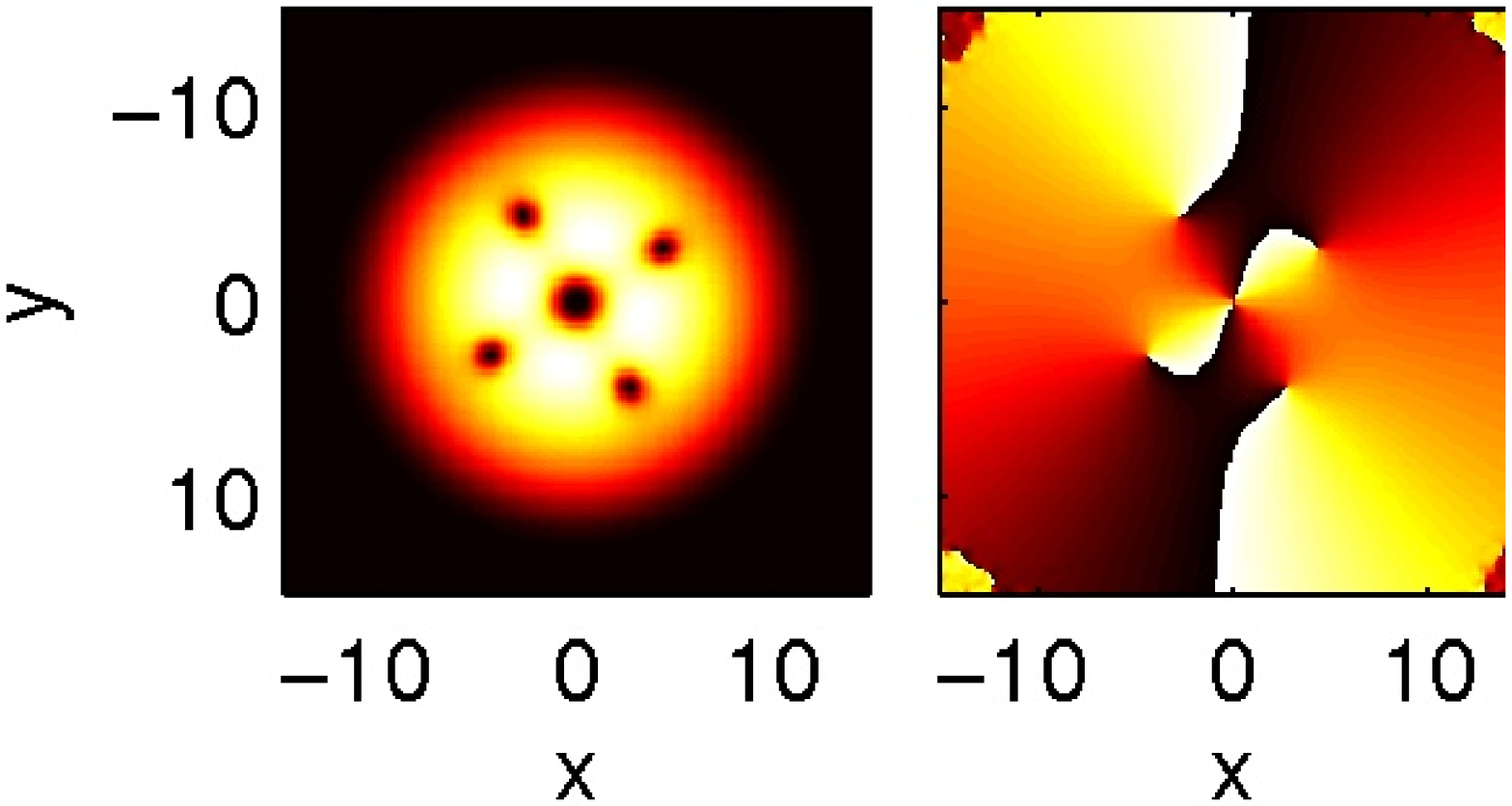} \\
&
\includegraphics[width=3.5cm,angle=0]{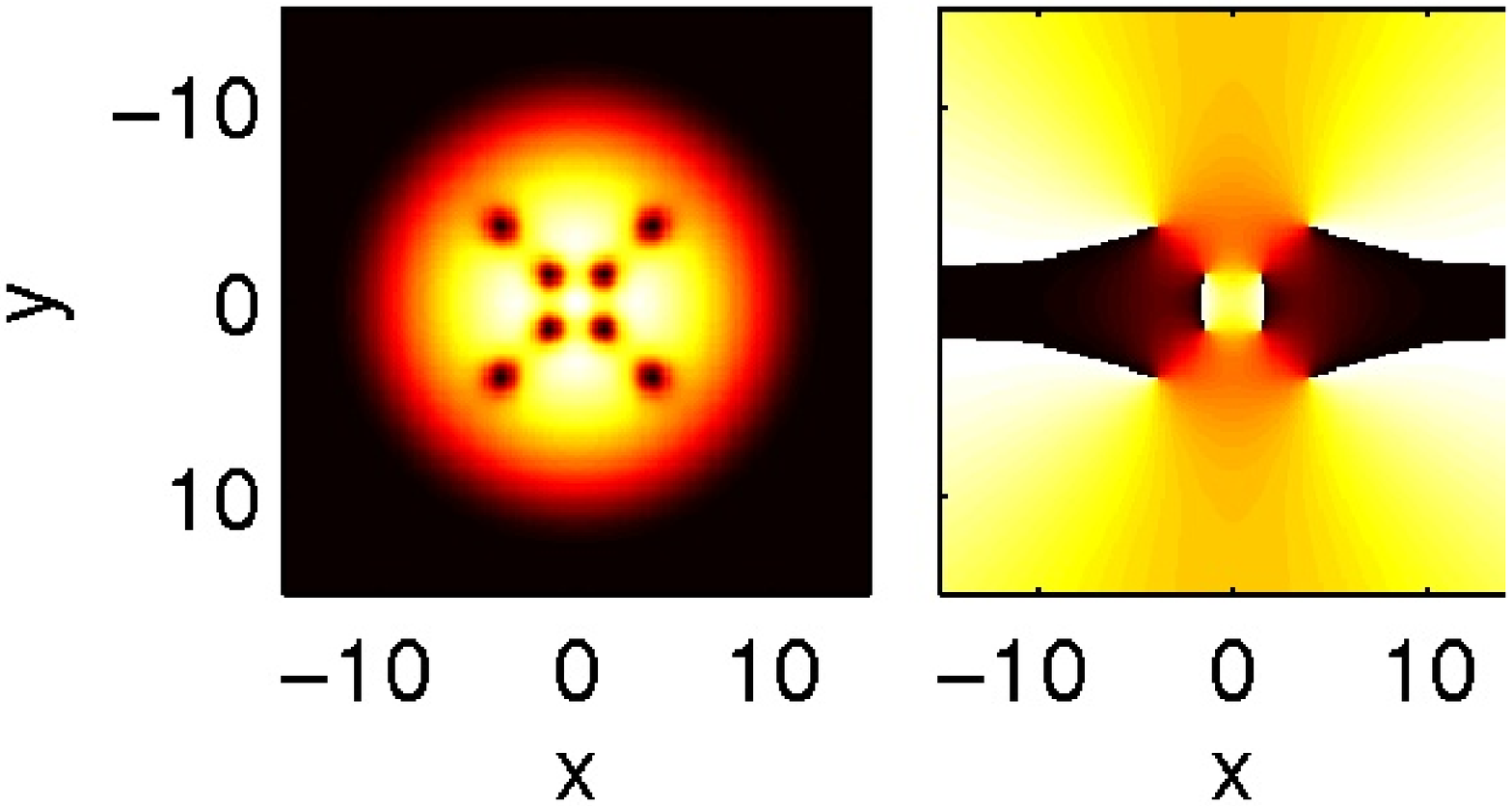}
\end{tabular}
\end{tabular}
\vspace{-0.2cm}
\caption{(Color online) 
Top left panel (a): Number of atoms as a function of the chemical
potential for the different states bifurcating from
the two dark soliton stripes (black lines) and
from the dark soliton cross 
(orange [gray in printed version] lines) for $\Omega=0.2$. 
Bottom left panel (b): Corresponding atom number difference with 
respect to the two dark soliton stripes branch. 
The occurrence of bifurcations at zero crossings of $\Delta N$
and the corresponding theoretical vertical line predictions thereof 
are similar
to the previous (and following) figures.
Middle column of panels (from top to bottom): 
Density and phase profiles (left and right subpanels respectively)
corresponding to bifurcating states from 
the two dark soliton stripes (top):
the aligned six- (6v2)
and 
eight-vortex state (8v2).
Right column of panels (from top to bottom): 
Density and phase profiles (left and right subpanels respectively)
corresponding to bifurcating states from 
the X-shaped dark soliton cross (top):
the diagonal six-vortex state (6x), 
the four and doubly-charged vortex state (5x),
and
the diagonal eight-vortex state (8x). 
}
\label{fig_2stripe_cross}
\end{figure*}

Numerical results on the symmetry-breaking bifurcations resulting
in the emergence of 
vortex cluster states from the first excited state (single dark soliton 
stripe) are summarized in Fig.~\ref{fig_DSS} for $\Omega=0.2$. The emergence
of $1 \times m$ ($m=2,3,4,5, \dots$) states can be observed to occur
respectively at 
$(0.68,0.98,1.26,1.54)$ 
while the corresponding theoretical predictions are 
$(\mu_{\rm cr}^{\rm vd},\mu_{\rm cr}^{\rm 3v},
\mu_{\rm cr}^{\rm 4v},\mu_{\rm cr}^{\rm 5v})=
(2/3,0.91,1.13,1.36)$. Clearly, the two-mode 
approach captures the fundamental phenomenology, although a slight progressive
degradation of the agreement on the critical point arises due to the
increasing departure from the linear limit. Additionally, the linear
stability properties of the resulting states directly reflect
the theoretical stability expectations discussed above; see also Ref.~\cite{vd}.

\begin{figure*}[htb]
\centering
\begin{tabular}{cc}
\begin{tabular}{c}
\includegraphics[width=4.5cm,angle=0]{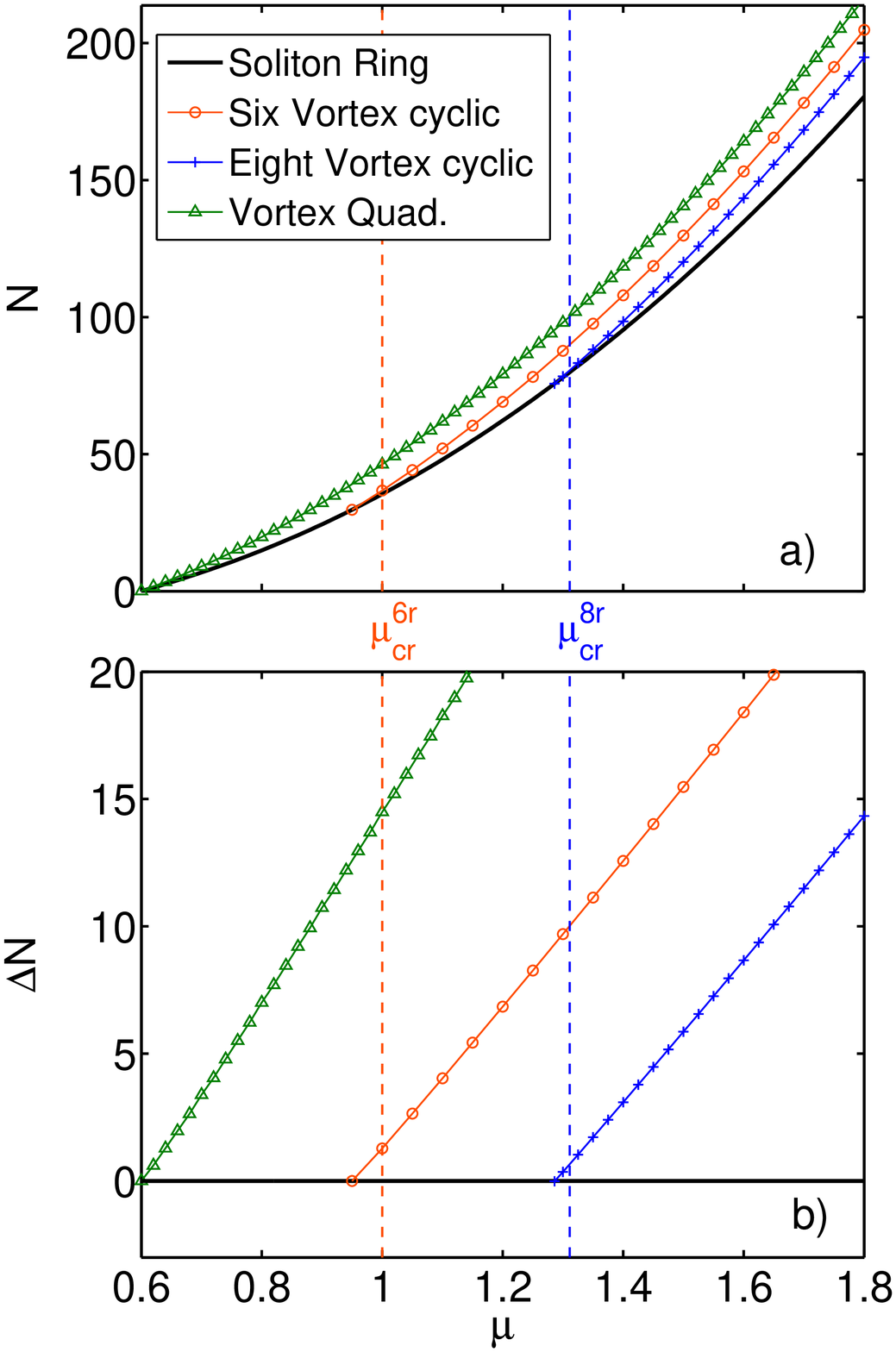}
\end{tabular}
&
\hskip0.5cm
\begin{tabular}{cc}
\includegraphics[width=3.5cm,angle=0]{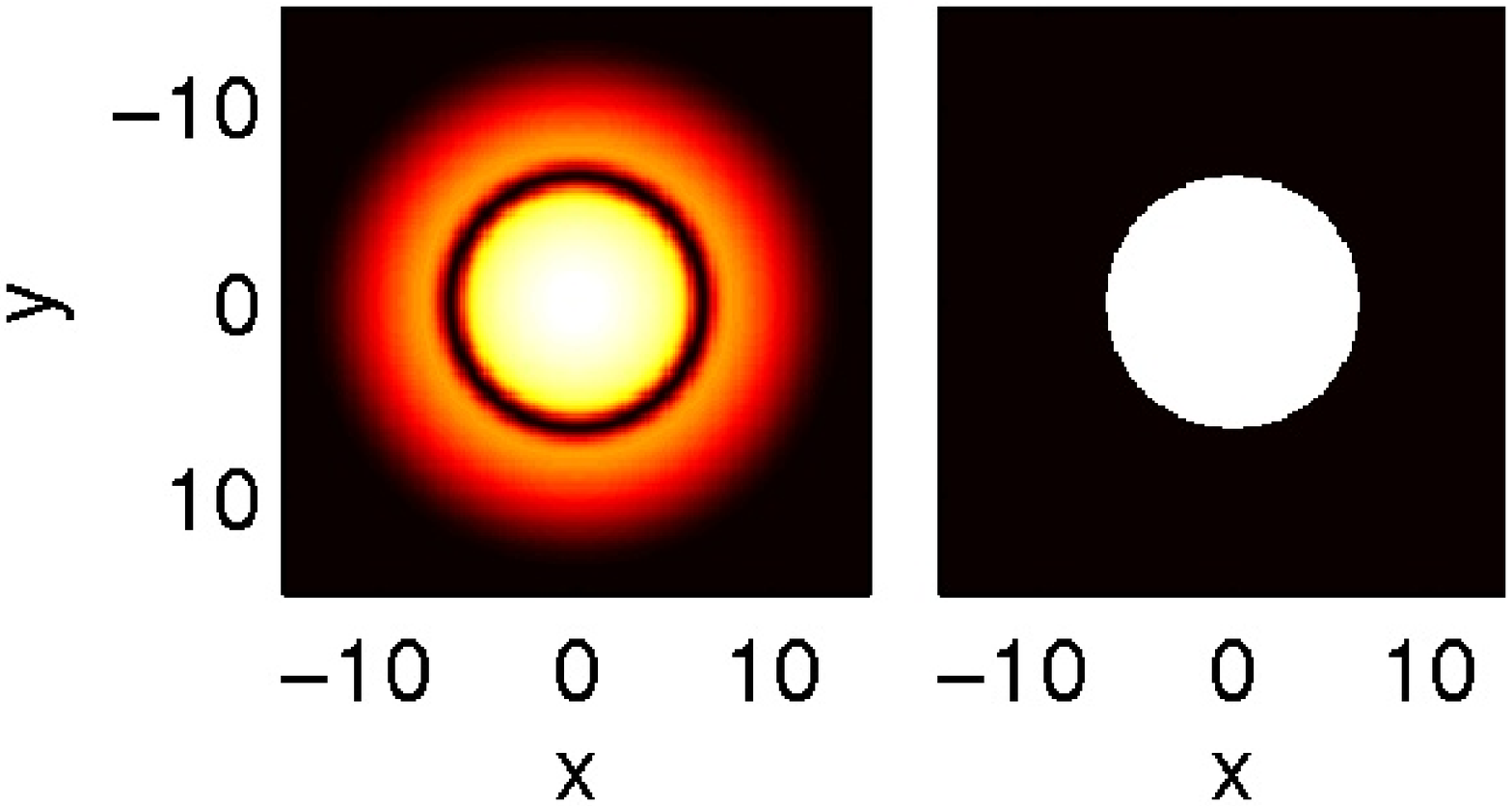}&
\includegraphics[width=3.5cm,angle=0]{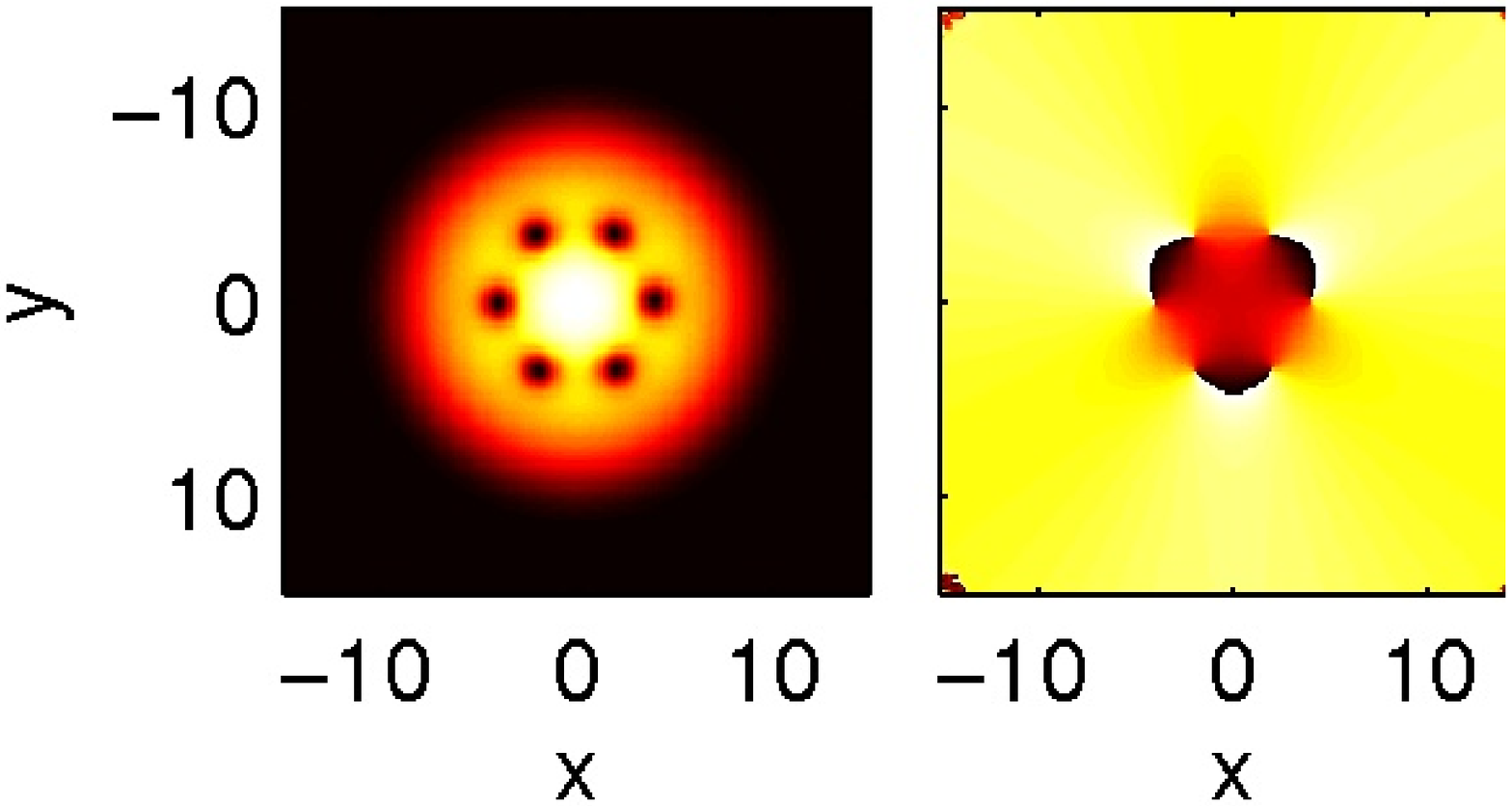} \\
\includegraphics[width=3.5cm,angle=0]{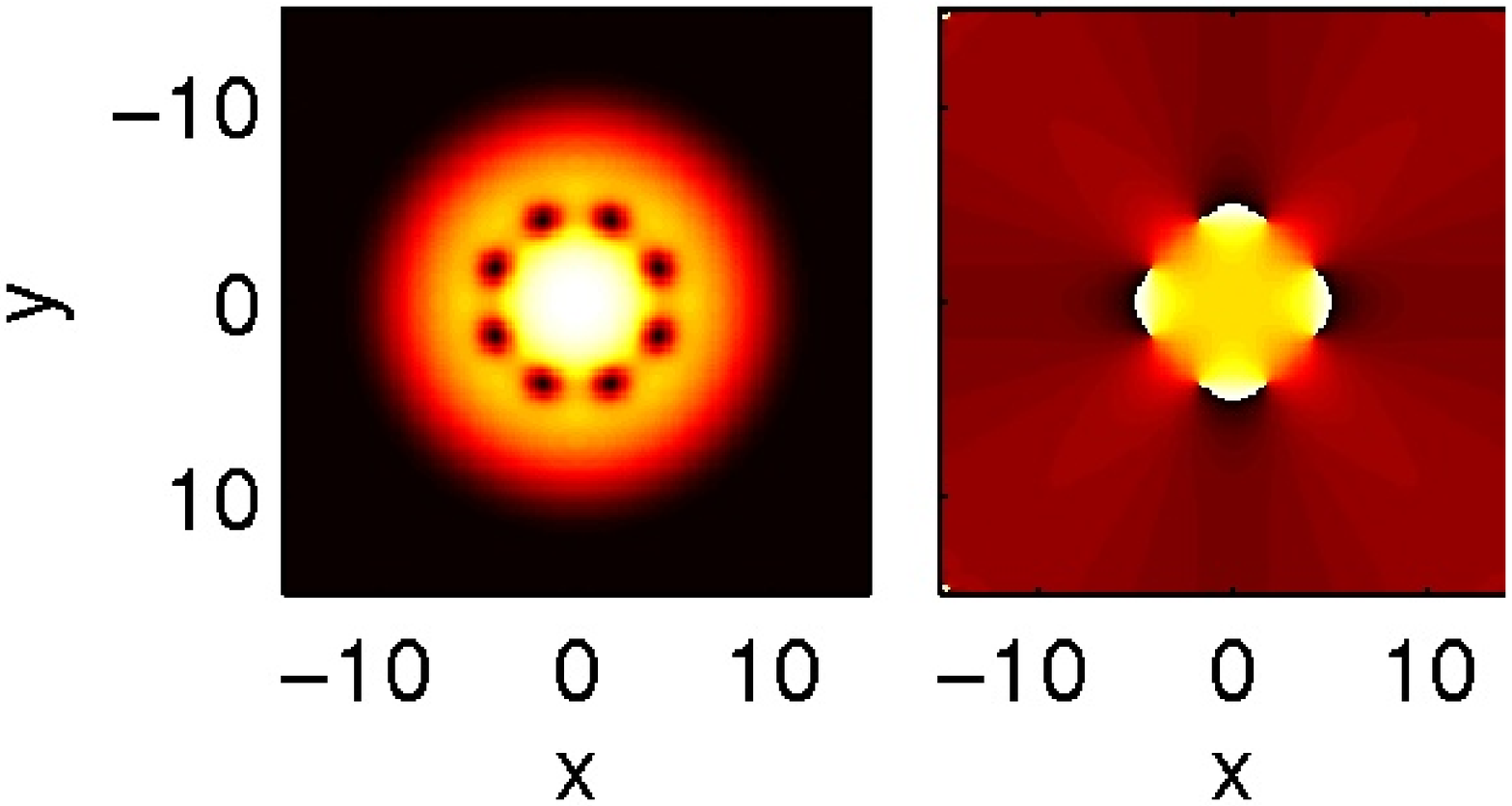} &
\includegraphics[width=3.5cm,angle=0]{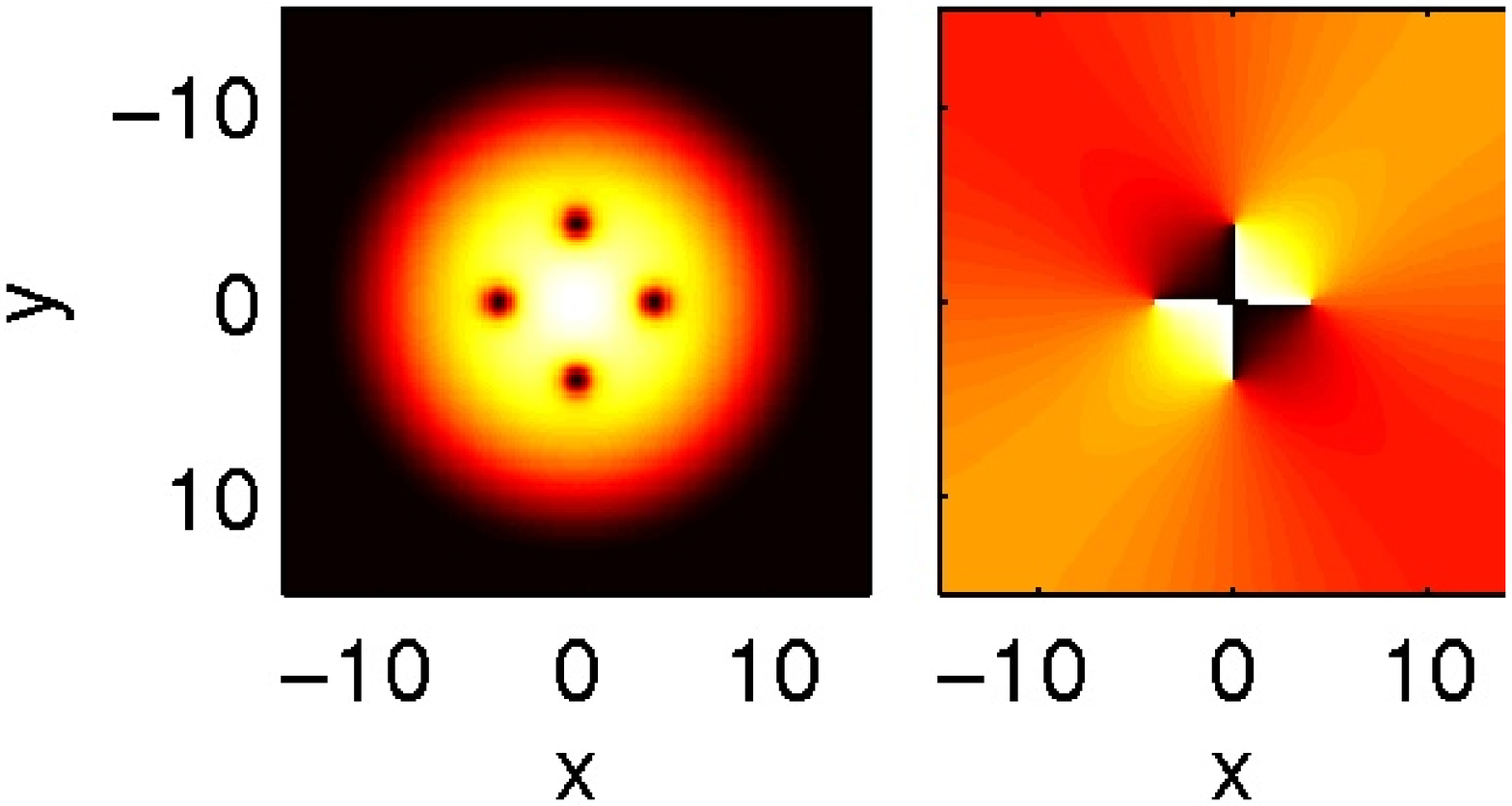} 
\end{tabular}
\end{tabular}
\vspace{-0.2cm}
\caption{(Color online) 
Top left panel (a): Number of atoms as a function of the chemical
potential for the different states bifurcating
from the ring dark soliton for $\Omega=0.2$. 
Bottom left panel (b): Corresponding atom number difference with 
respect to the ring dark soliton.
Right panels (from left to right and top to bottom): 
Density and phase profiles (left and right subpanels respectively)
corresponding to bifurcating states from 
the ring dark soliton state (top left):
%
the hexagonal (6r) and
octagonal (8r) vortex states. 
Bottom right panel: 
vortex quadrupole state (note that this state does {\em not}
bifurcate from the ring dark soliton).
}
\label{fig_ring}
\end{figure*}

Importantly, this approach is not restricted to the first
excited state. The advantage of the bifurcation method and of
the wealth of states that can be derived from it is unveiled, 
e.g., when considering the next set of excited states, namely 
the combinations with $n+m=2$, i.e., the 
degenerate states $u_{20}$, $u_{02}$ and $u_{11}$. In this
setting, already a vortex {\it quadrupole} can be formed
at the linear limit as $u_{20}+i u_{02}$ \cite{todd}. Interestingly, so
can a doubly-charged vortex through $u_{20} - u_{02} + 2 i u_{11}$.
However, these states do not present symmetry breaking bifurcations
and, therefore, are not considered further in what follows.
Focusing on the states that do, some 
prototypical examples are 
(i) the solitonic state consisting of two stripes 
(i.e., a two-dark-soliton state), i.e., the $u_{20}$ state,
(ii) an X-shaped dark soliton cross emerging from $u_{20}-u_{02}$, as well as
(iii) a ring dark soliton state \cite{ring1} (see also Refs.~\cite{fr4,herring}) 
arising from $u_{20} + u_{02}$.
The $u_{20}$ state is one for which the generalization of the
phenomenology of Fig.~\ref{fig_DSS} is most
straightforward as with $\phi_0=u_{20}$ and $\phi_1=u_{0m}$
states with $2 \times m$ (2 lines of $m$ vortices each) are
formed with $m>2$. 
For example, for $\phi_1=u_{03}$ and $\phi_1=u_{04}$, such
bifurcations are predicted at 
$\mu_{\rm cr}^{\rm 6v2}=283 \Omega/67$, and 
$\mu_{\rm cr}^{\rm 8v2}=2965 \Omega/551$
(see thick black vertical dashed lines in Fig.~\ref{fig_2stripe_cross}),
respectively, leading to $6=2 \times 3$ and $8=2 \times 4$
two-line vortex clusters. 
On the other hand,
the X-shaped dark soliton pair is subject to 
similar symmetry-breaking bifurcations/destabilizations due
to $\phi_1=u_{03}$, $u_{13}$, $u_{04}$ etc. These bifurcations, 
occurring in turn at 
$\mu_{\rm cr}^{\rm 6x}=33 \Omega/7$,
$\mu_{\rm cr}^{\rm 5x}=99 \Omega/17$, and
$\mu_{\rm cr}^{\rm 8x}=369 \Omega/59$ (see thin orange 
[gray in printed version] vertical dashed lines 
in Fig.~\ref{fig_2stripe_cross}), lead to 
a diagonal state with 6 vortices, 
a doubly-charged vortex in the center together with 
a four-vortex quadrupole around it, and 
an 8-vortex cluster of near-diagonal vortices. 
Finally, the ring dark soliton
gets mixed with states of the form  $\phi_1=q_l(r)\sin(l\theta)$, where
$q_l(r)=\sqrt{\frac{2}{2!\pi}}(r^l)\exp(-\Omega r^2/2)$, again for
$l>3$. Remarkably, these symmetry-breaking events, which can be predicted
to occur, e.g., at 
$\mu_{\rm cr}^{\rm 6r}=5 \Omega$ and
$\mu_{\rm cr}^{\rm 8r}=59 \Omega/9$
(see, respectively, orange [gray in printed version] and blue [dark in
printed version] vertical dashed lines in Fig.~\ref{fig_ring}), 
for $l=3$ and $4$, give
rise to polygonal vortex configurations with the vortices now placed
on the periphery of the circle. This way, vortex hexagons, octagons,
decagons, etc. can be systematically constructed at will.

\begin{figure}[htb]
\centering
\begin{tabular}{cc}
\includegraphics[width=4.4cm,height=4.0cm,angle=0]{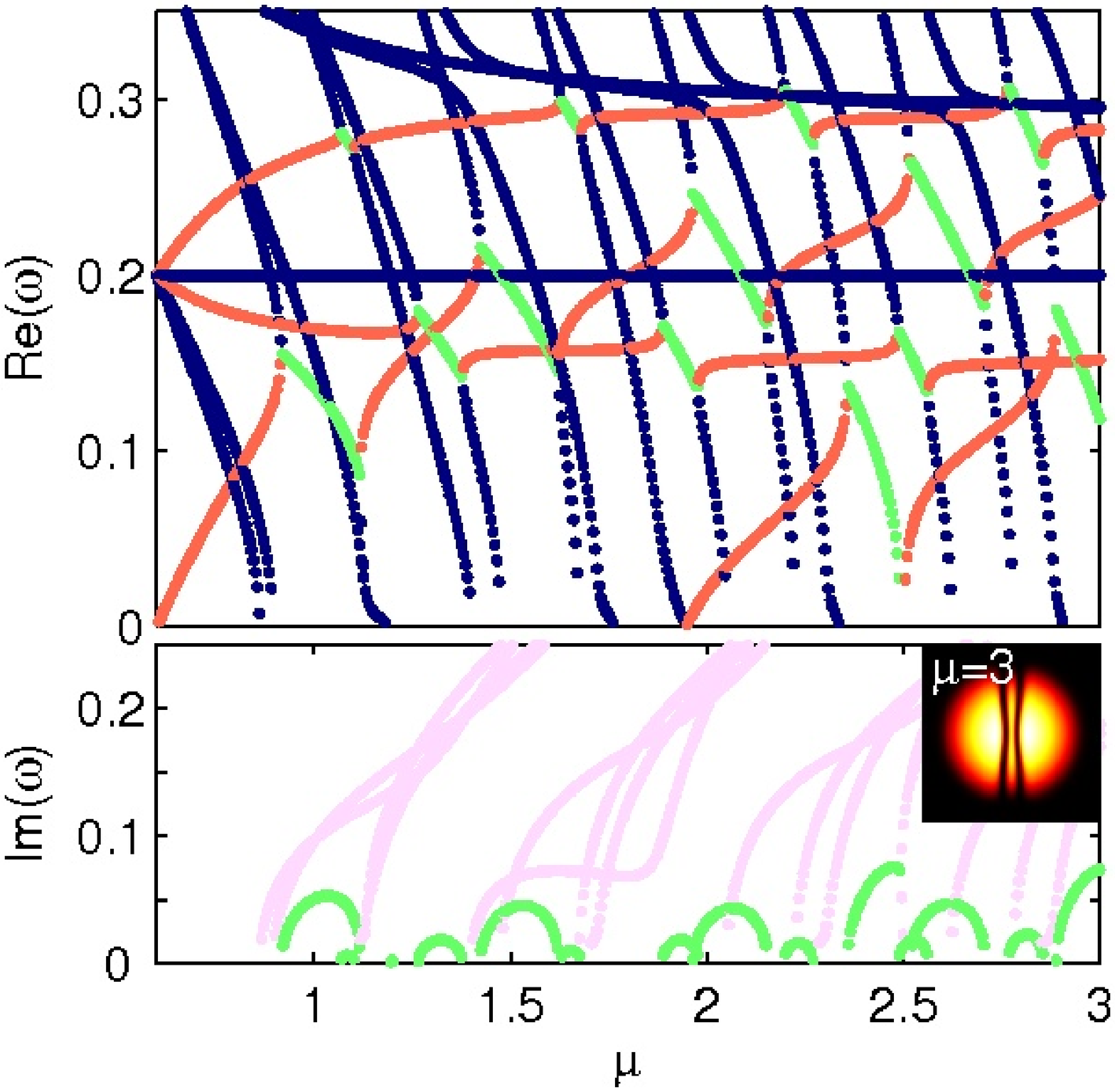} &
\includegraphics[width=4.4cm,height=4.0cm,angle=0]{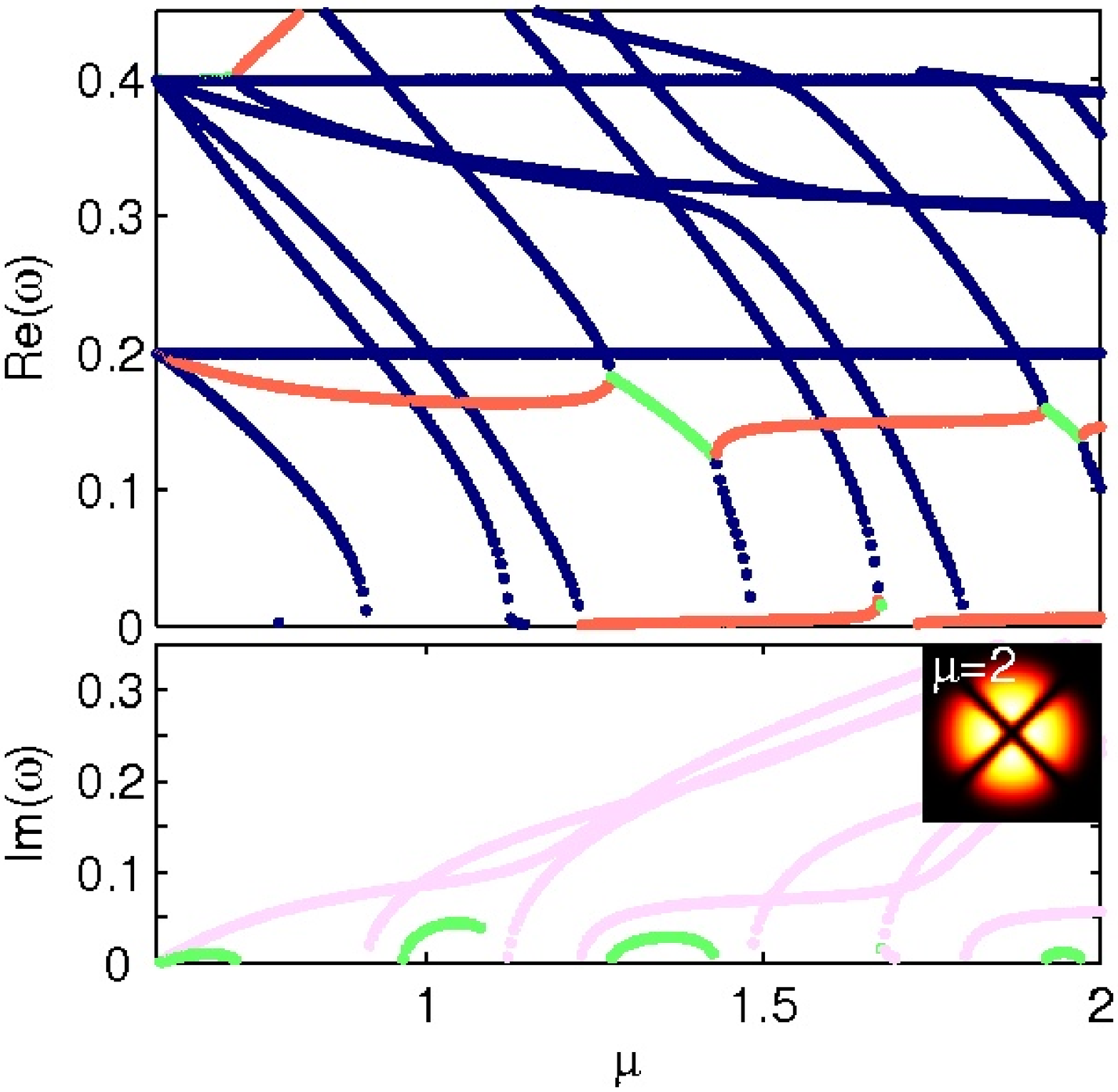} \\[-0.1cm]
\includegraphics[width=4.4cm,height=4.0cm,angle=0]{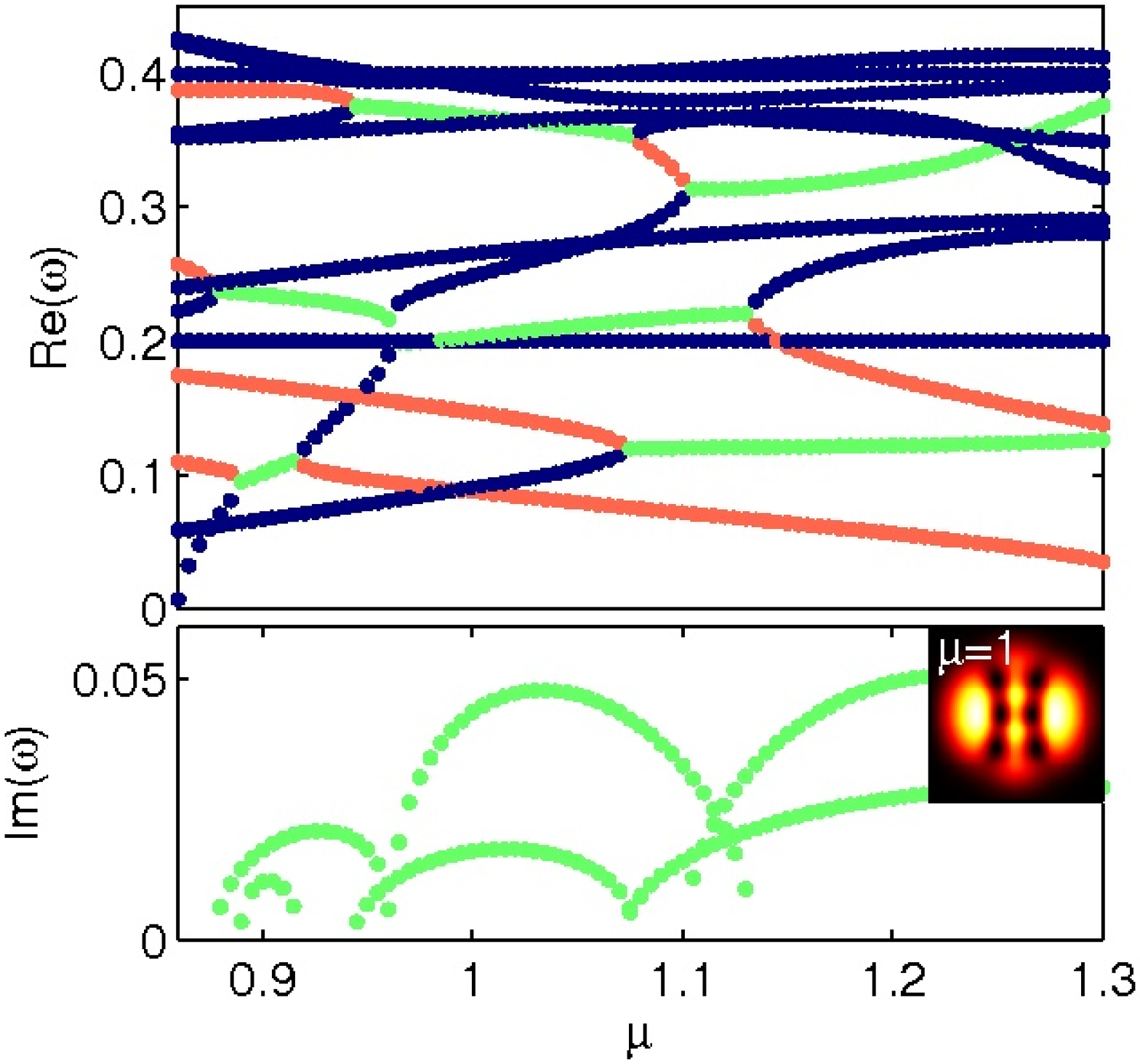} &
\includegraphics[width=4.4cm,height=4.0cm,angle=0]{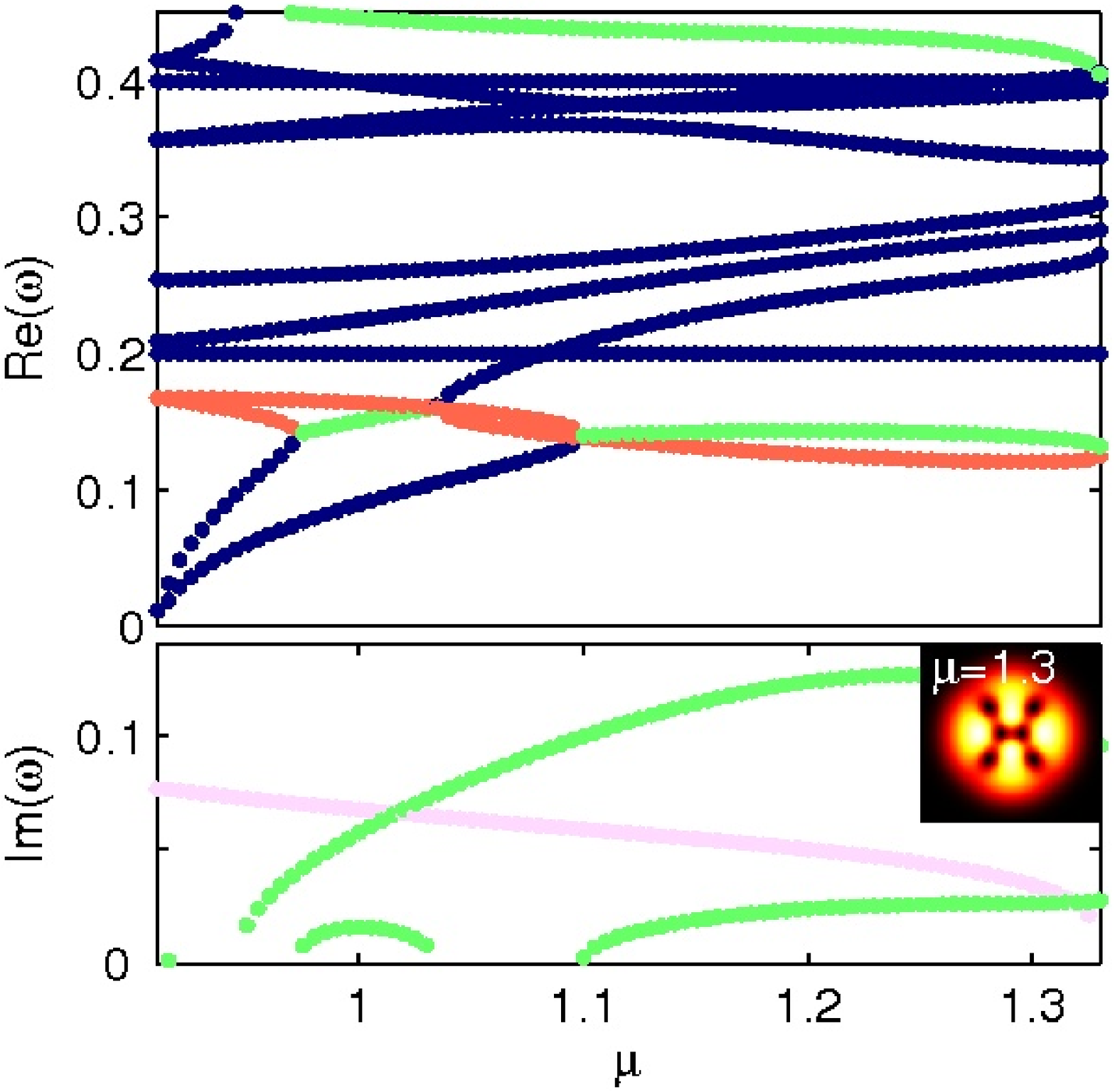} \\[-0.1cm]
\includegraphics[width=4.4cm,height=4.0cm,angle=0]{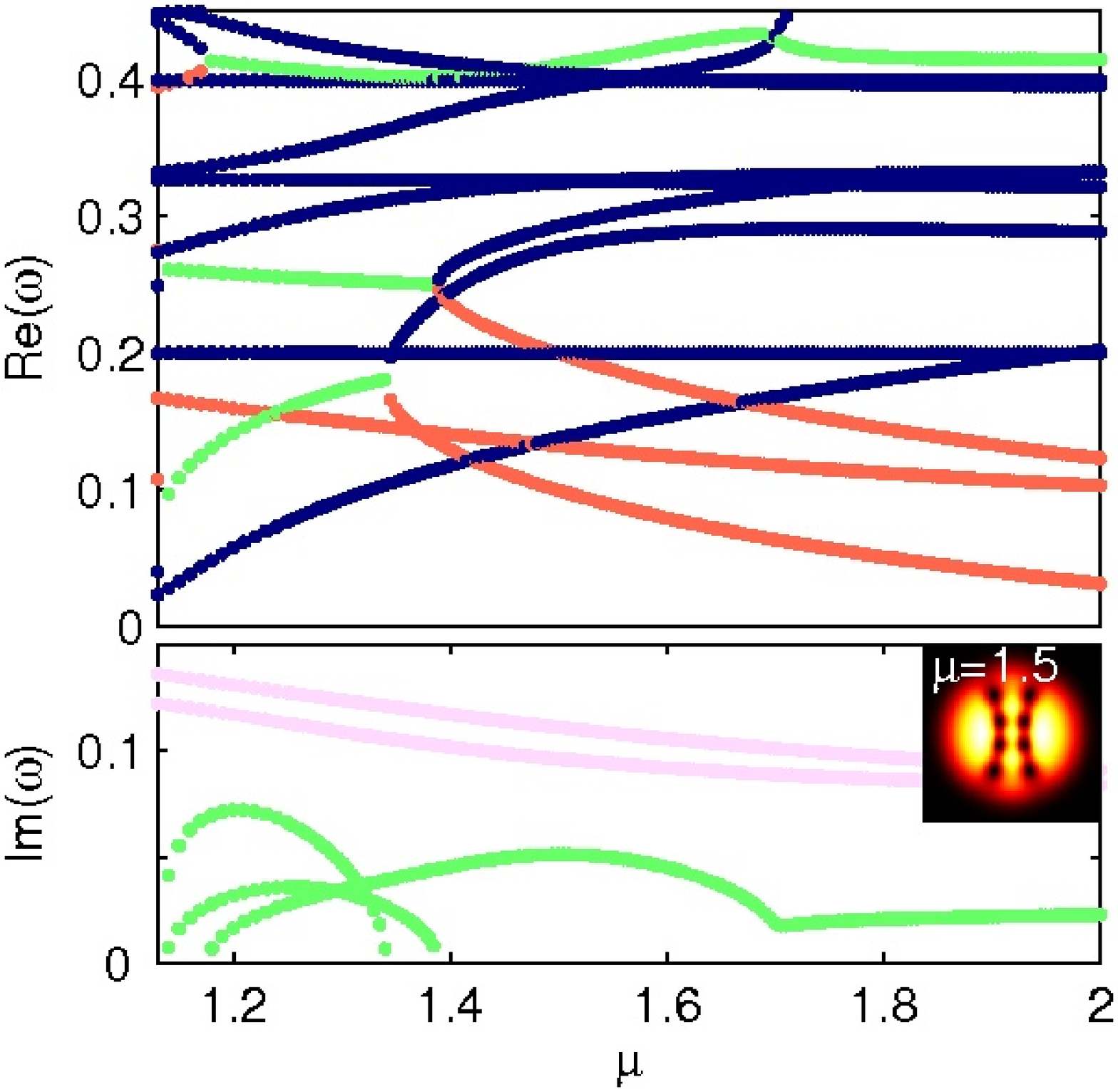}&
\includegraphics[width=4.4cm,height=4.0cm,angle=0]{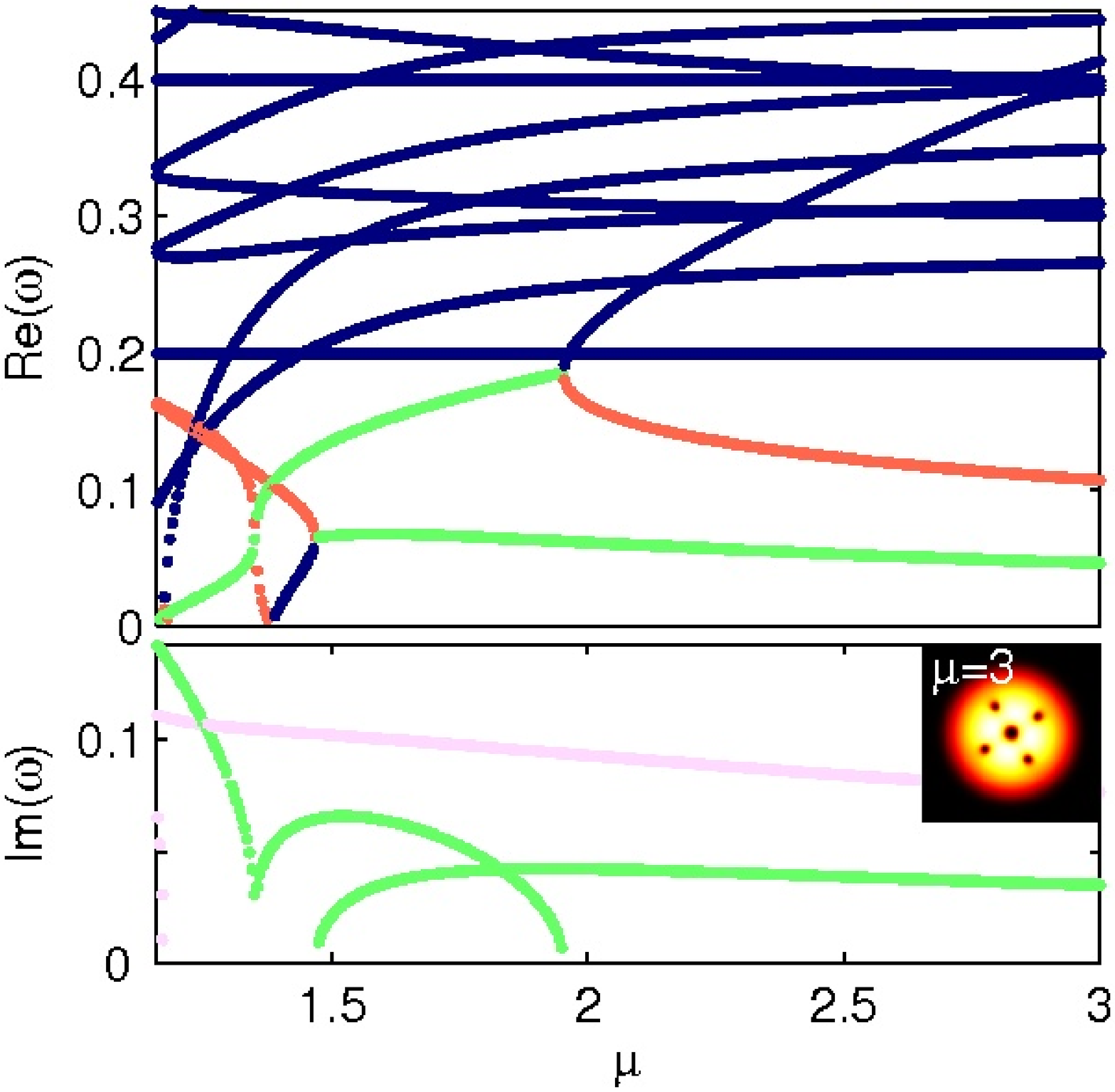} \\[-0.1cm]
&
\includegraphics[width=4.4cm,height=4.0cm,angle=0]{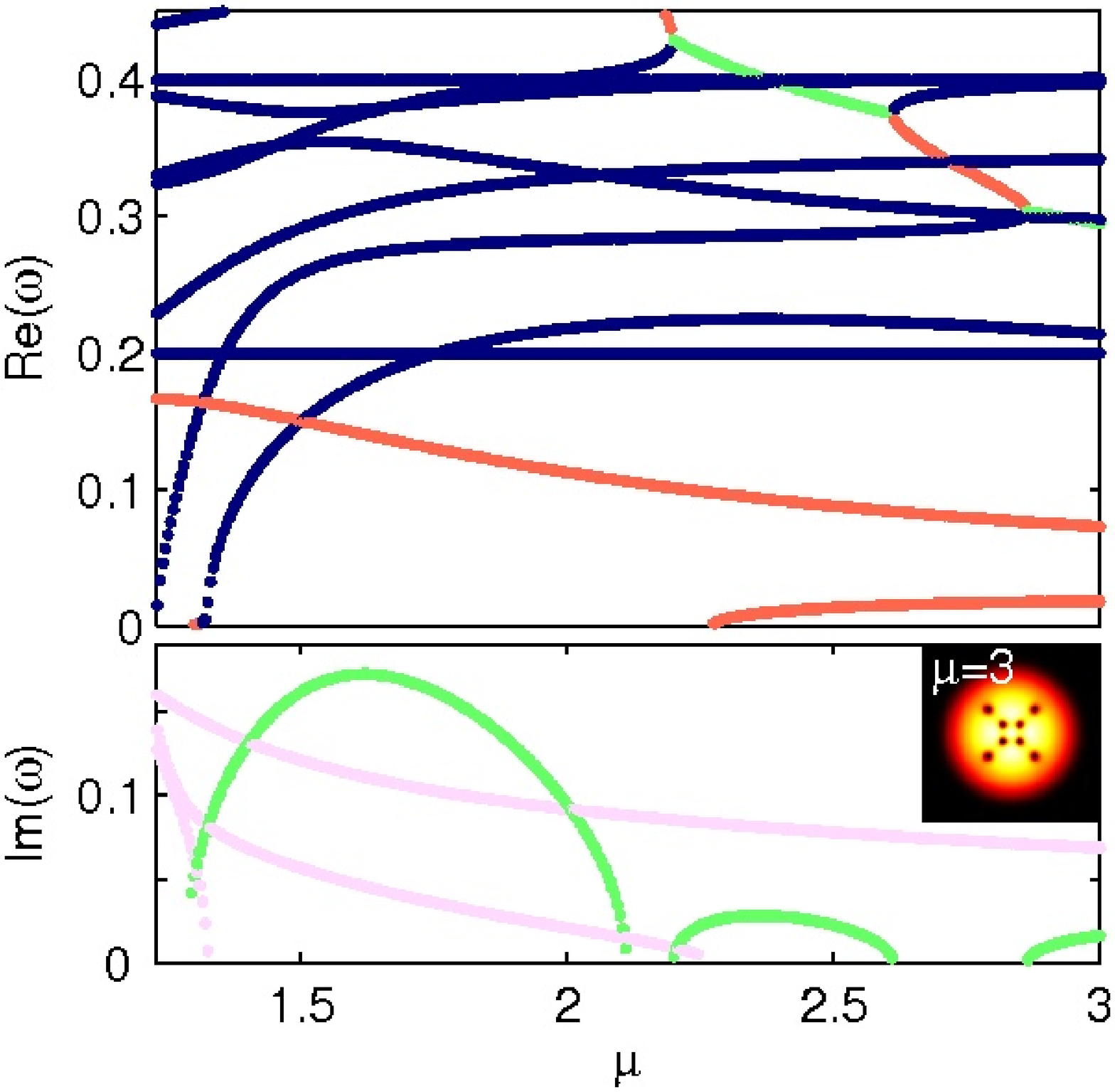}
\end{tabular}
\vspace{-0.5cm}
\caption{(Color online) 
BdG spectra for the two-dark-soliton stripe and its
bifurcating states (left column) and for the dark soliton
cross and its bifurcating states (right column)
corresponding to the states described in Fig.~\ref{fig_2stripe_cross}.
The inset in each of the spectra depicts a typical configuration
at the chemical potential indicated.
In all BdG spectra we depict the real (top subpanel) and imaginary 
(bottom subpanel) parts of the eigenfrequencies as a function of the
chemical potential $\mu$ using the following color coding for
the online (respectively, in print) version:
blue (black)=positive energy (Krein sign) modes,
orange (dark gray)=negative energy modes,
green (gray)=oscillatory unstable modes, and
pink (light gray)=non-oscillatory unstable modes.
The presence of non-vanishing imaginary parts is an indication of instability. 
Note that the only stable solutions are the 
two-dark soliton stripe (top left panel) and its first
symmetry breaking offspring, the six-vortex state (left panel
of second row), which are
stable only for low enough chemical potentials.
%
}
\label{fig_BdG_2stripe_cross}
\end{figure}

\begin{figure}[htb]
\centering
\begin{tabular}{cc}
\includegraphics[width=4.4cm,height=4.0cm,angle=0]{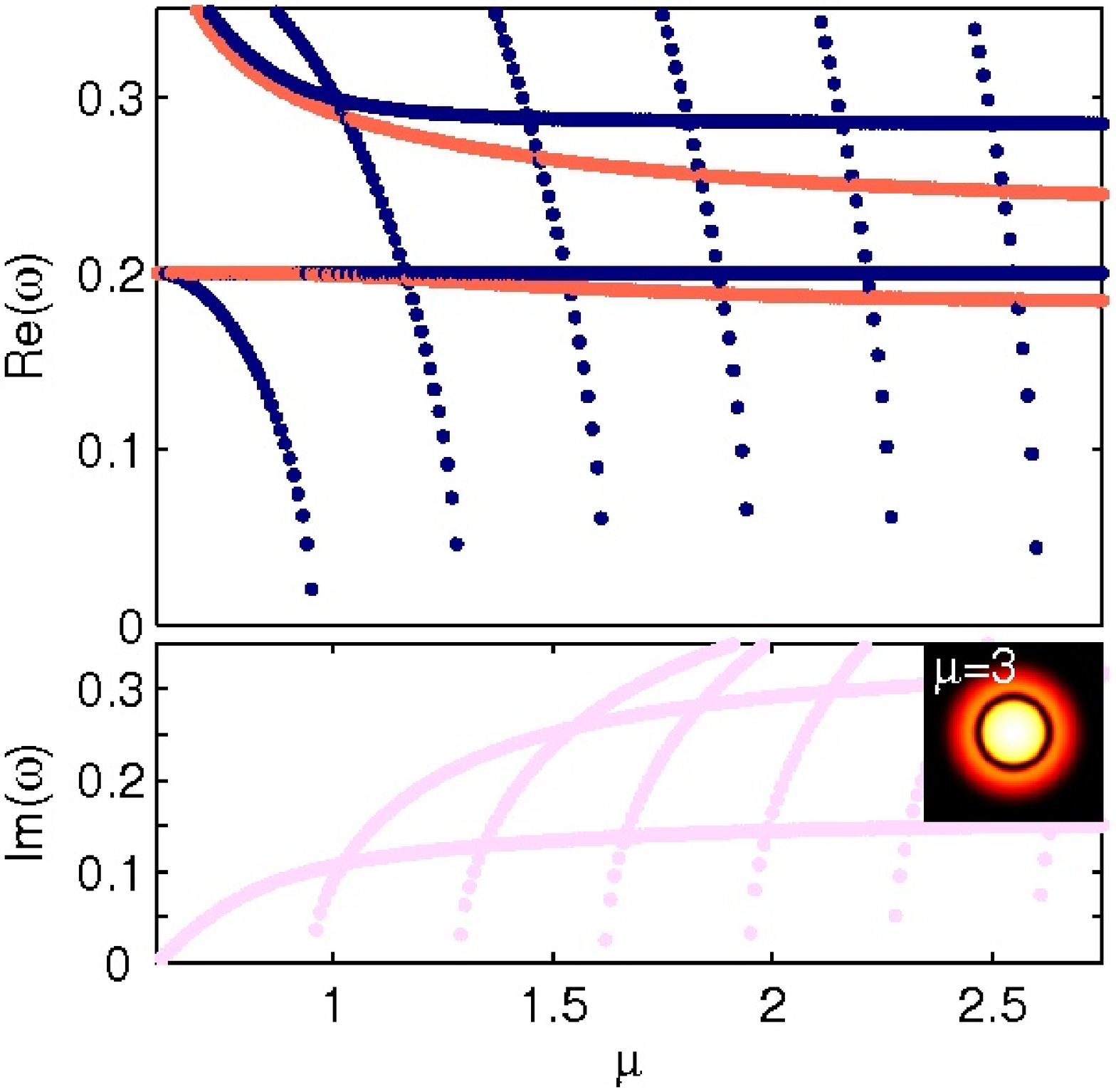} &
\includegraphics[width=4.4cm,height=4.0cm,angle=0]{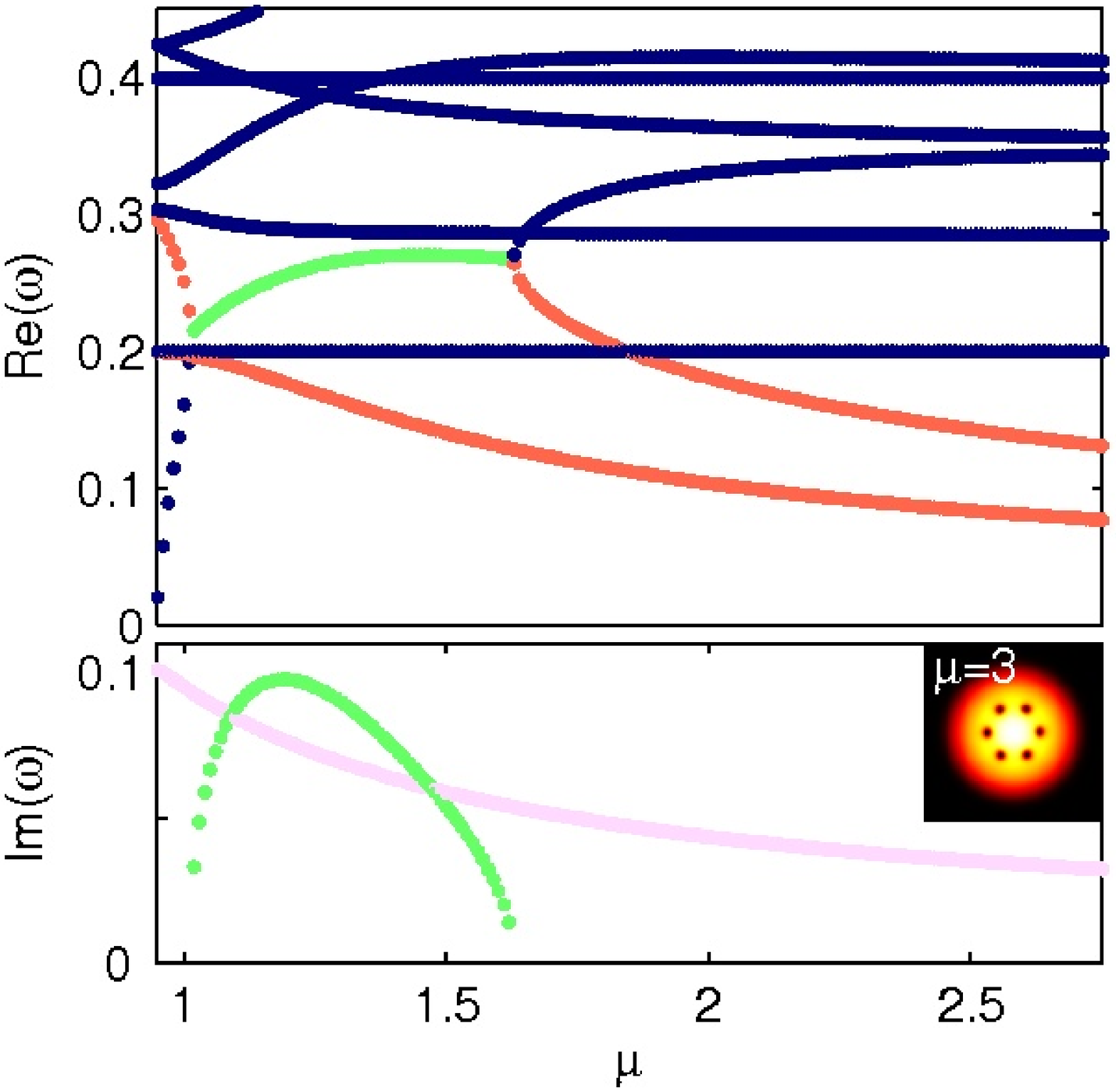} \\[-0.1cm]
\includegraphics[width=4.4cm,height=4.0cm,angle=0]{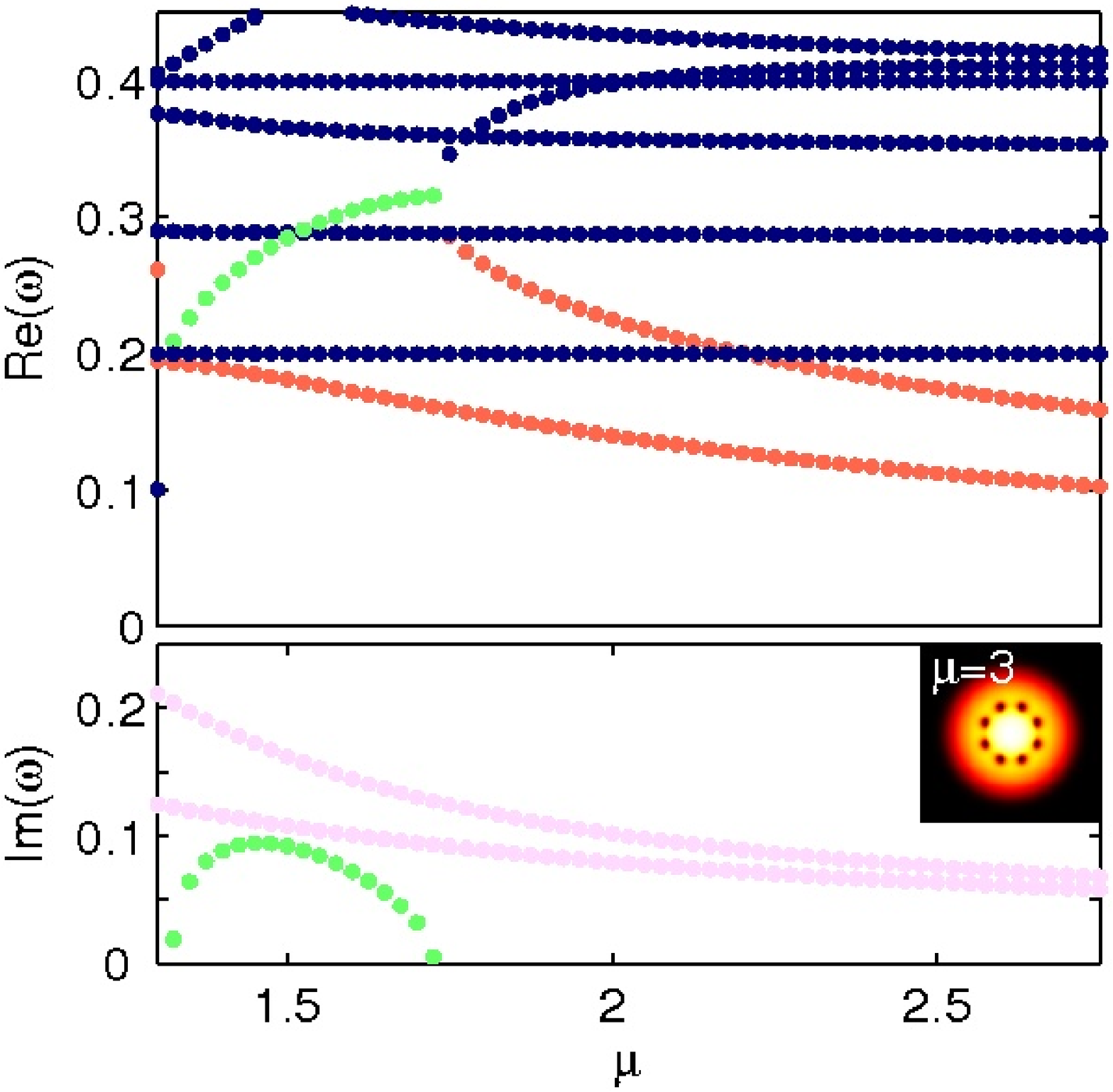} &
\includegraphics[width=4.4cm,height=4.0cm,angle=0]{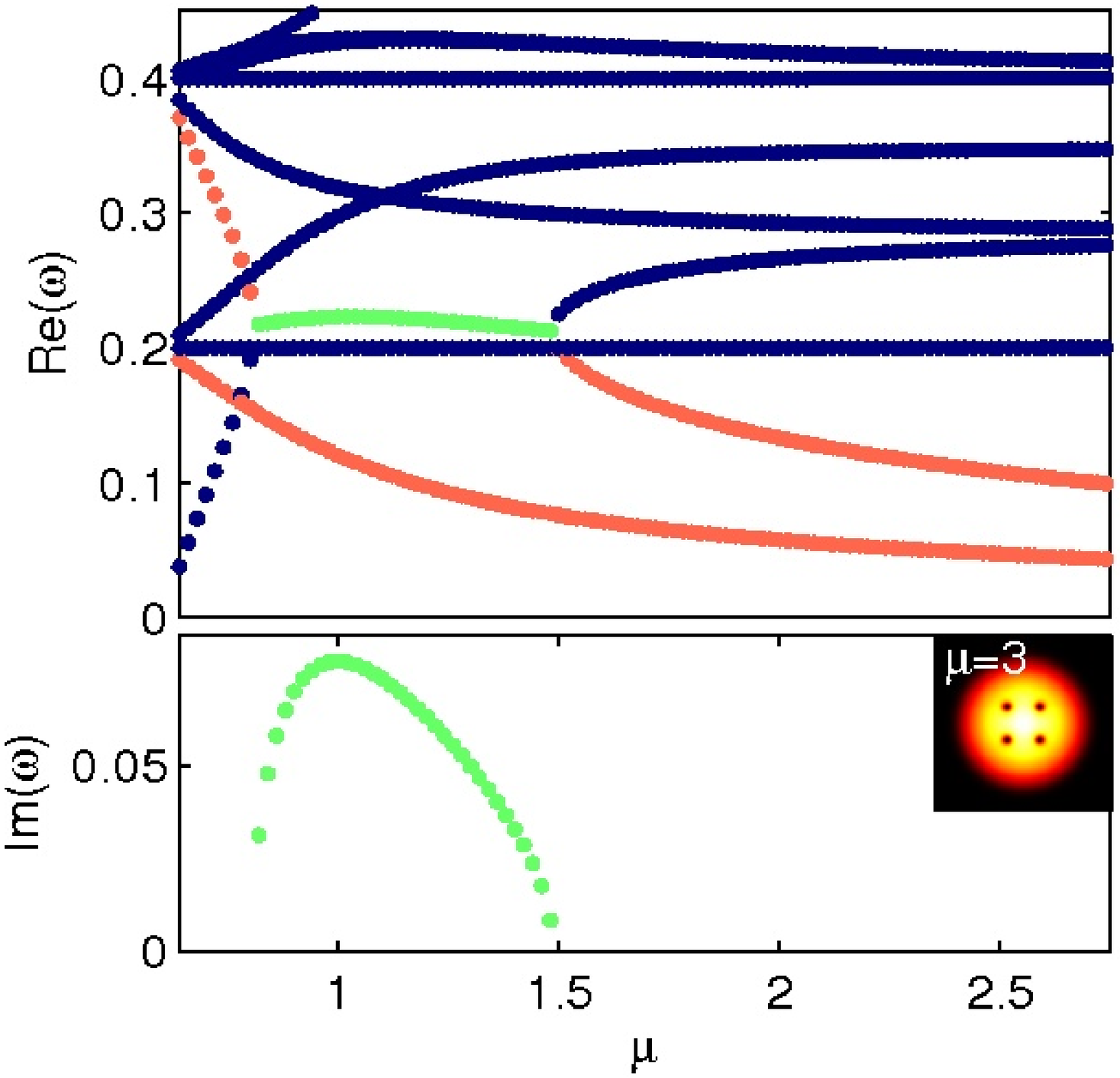}
\end{tabular}
\vspace{-0.5cm}
\caption{(Color online) 
BdG spectra for the states bifurcating from the vortex ring and for the
vortex quadrupole (bottom right) that does not bifurcate from the
vortex ring corresponding to the bifurcations depicted in
Fig.~\ref{fig_ring}.
For an explanation of the color codes see Fig.~\ref{fig_BdG_2stripe_cross}.
From the above it is clear that the most robust state is the vortex
quadrupole (which is dynamically stable for small $\mu$ except for a narrow 
interval of oscillatory instabilities).
}
\label{fig_BdG_ring}
\end{figure}

\begin{figure*}[htb]
\centering
\begin{tabular}{cc}
\hskip-0.5cm
\begin{tabular}{c}
\includegraphics[width=4.5cm,angle=0]{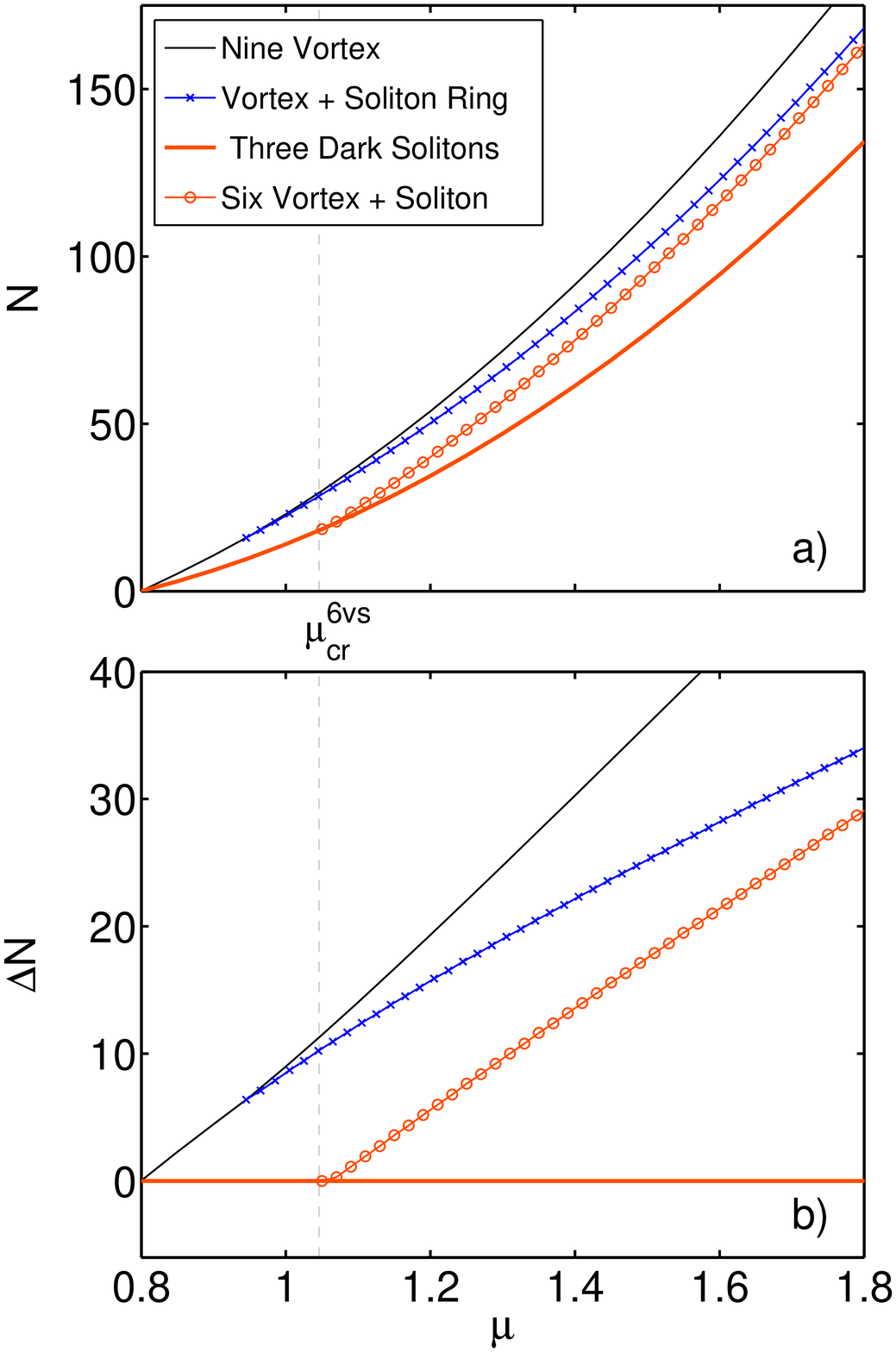}
\end{tabular}
&
\hskip0.5cm
\begin{tabular}{cc}
\includegraphics[width=3.5cm,angle=0]{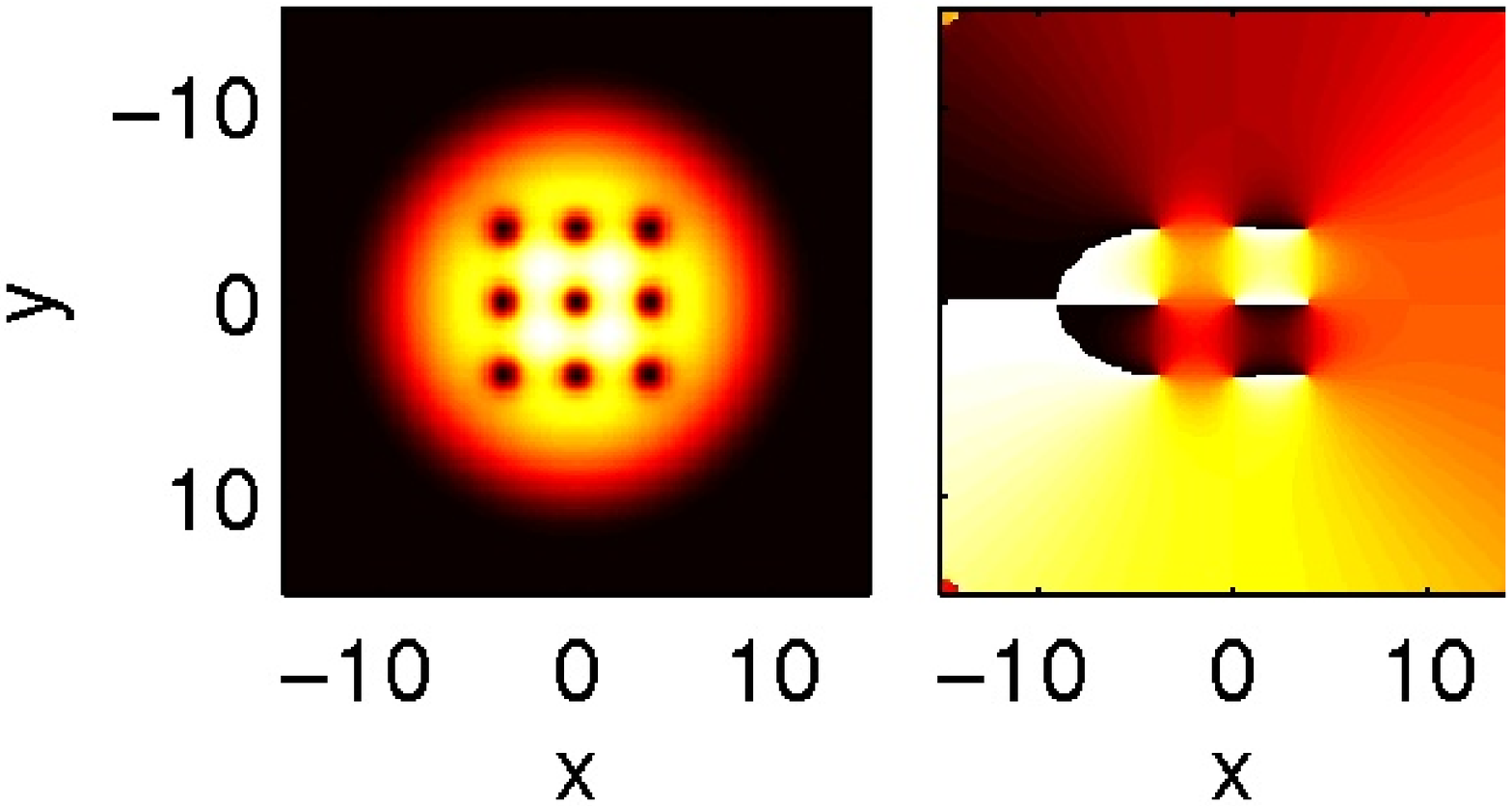} &
\includegraphics[width=3.5cm,angle=0]{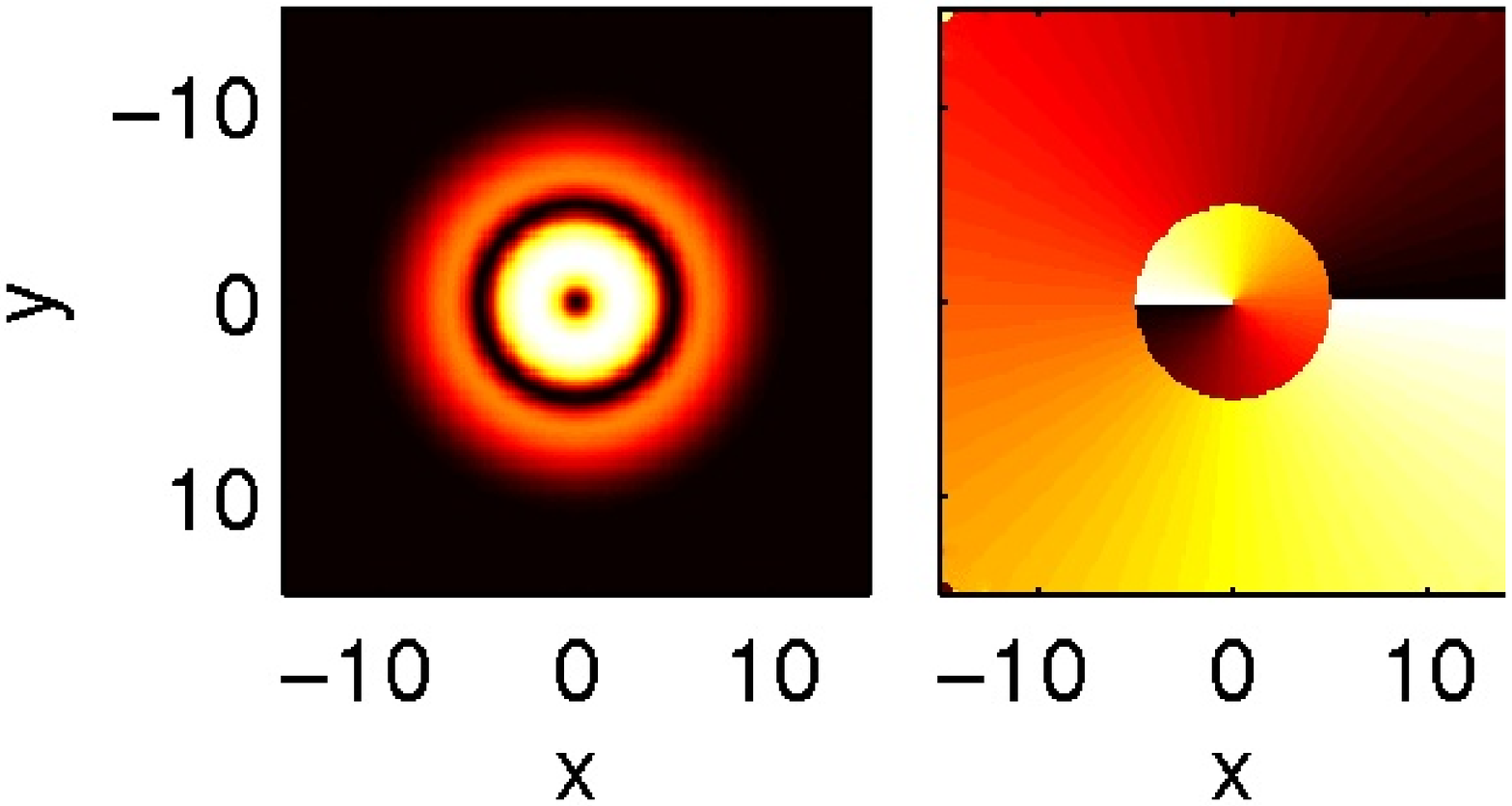} \\
\includegraphics[width=3.5cm,angle=0]{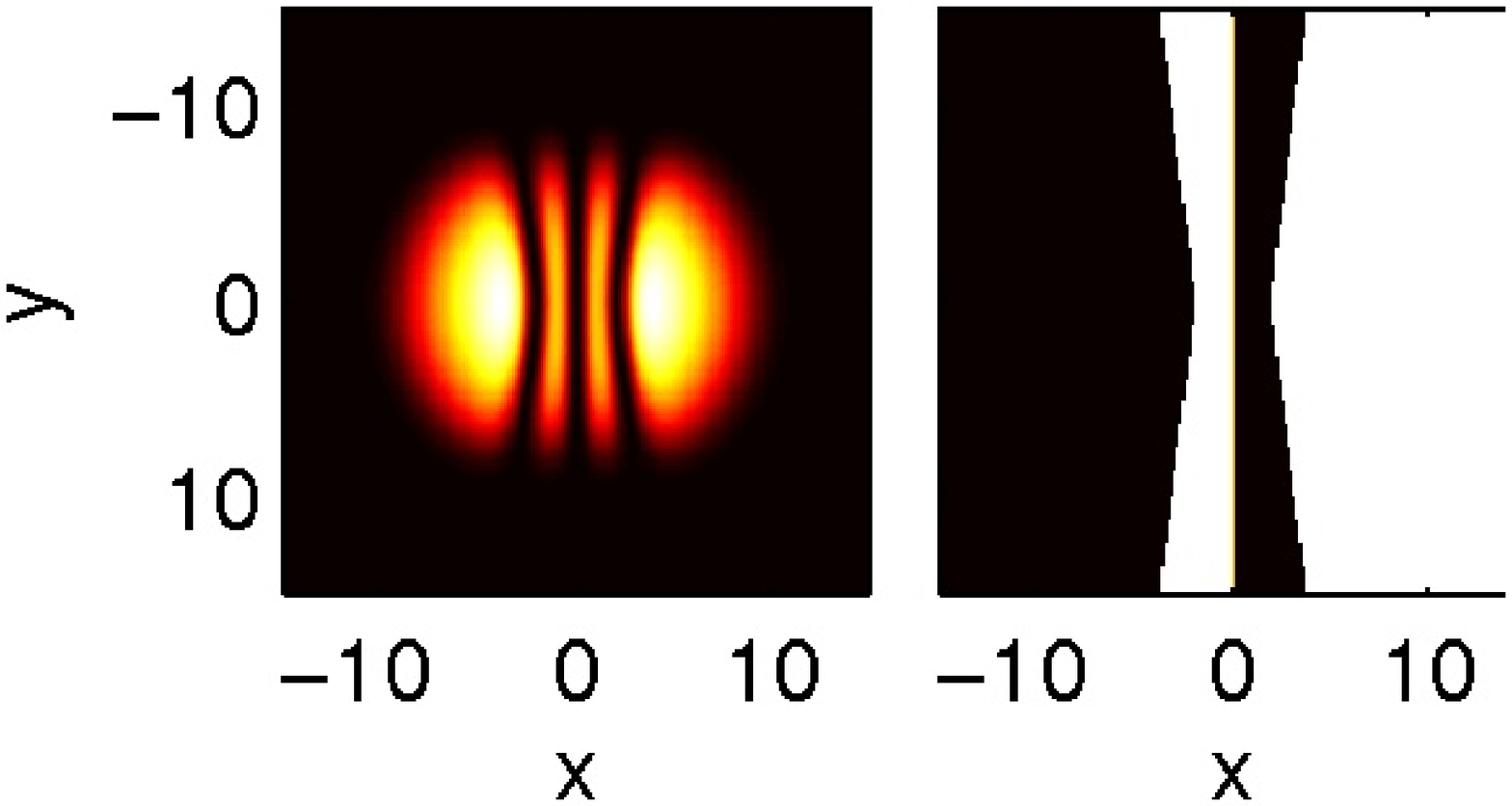} &
\includegraphics[width=3.5cm,angle=0]{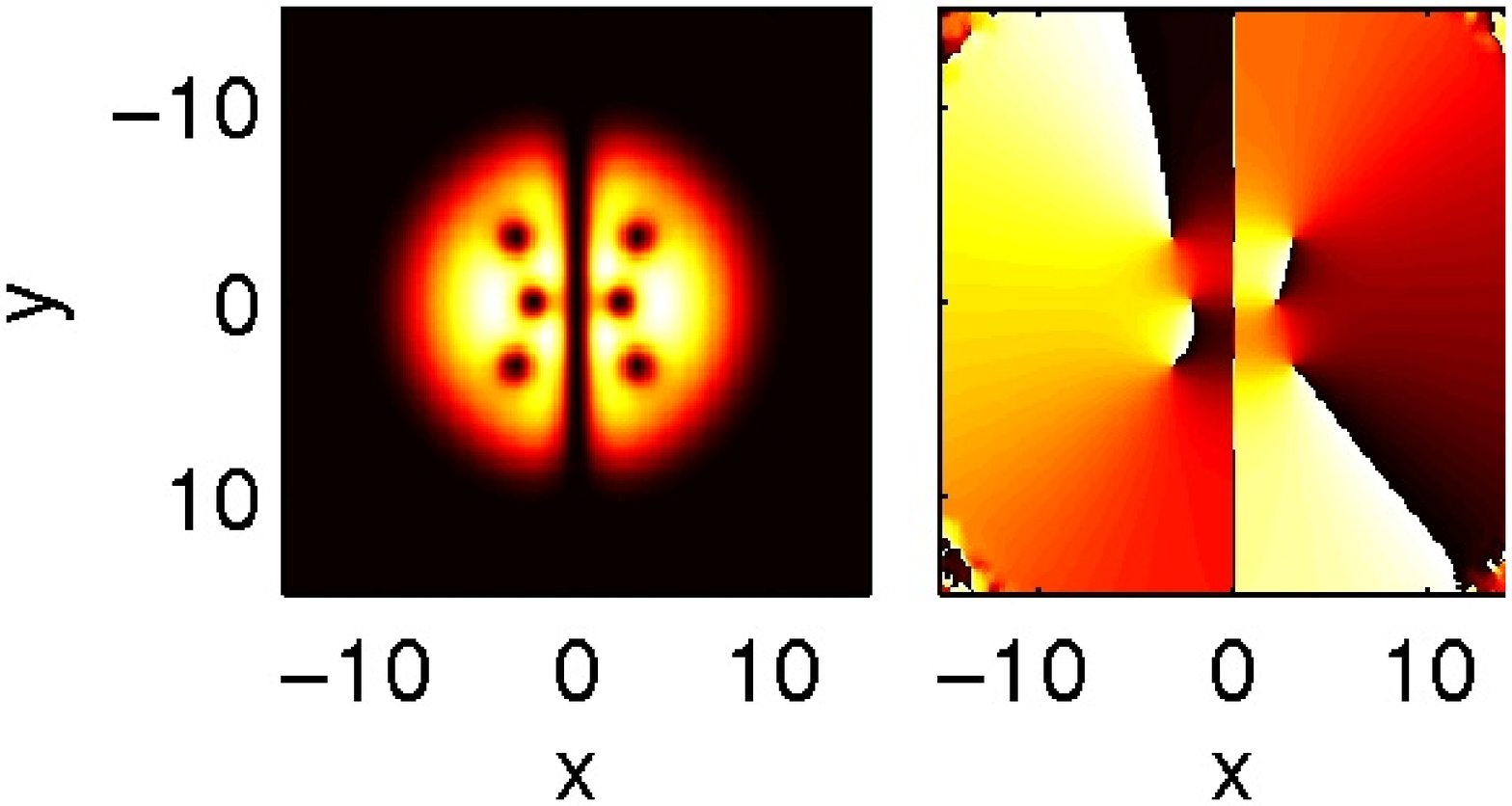}
\end{tabular}
\end{tabular}
\vspace{-0.2cm}
\caption{(Color online) 
Top left panel (a): Number of atoms as a function of the chemical
potential for the different states bifurcating
from $\mu=4 \Omega$ for $\Omega=0.2$. 
Bottom right panel (b): Corresponding atom difference with respect to the three-soliton state. 
Right panels (from left to right and top to bottom):
Density and phase profiles of the nine-vortex state, the vortex-soliton-ring, the three-soliton state and the mixed state 
consisting of a dark stripe and six vortices.
}
\label{fig_other}
\end{figure*}

\begin{figure}[htb]
\centering
\begin{tabular}{cc}
\includegraphics[width=4.4cm,height=4.0cm,angle=0]{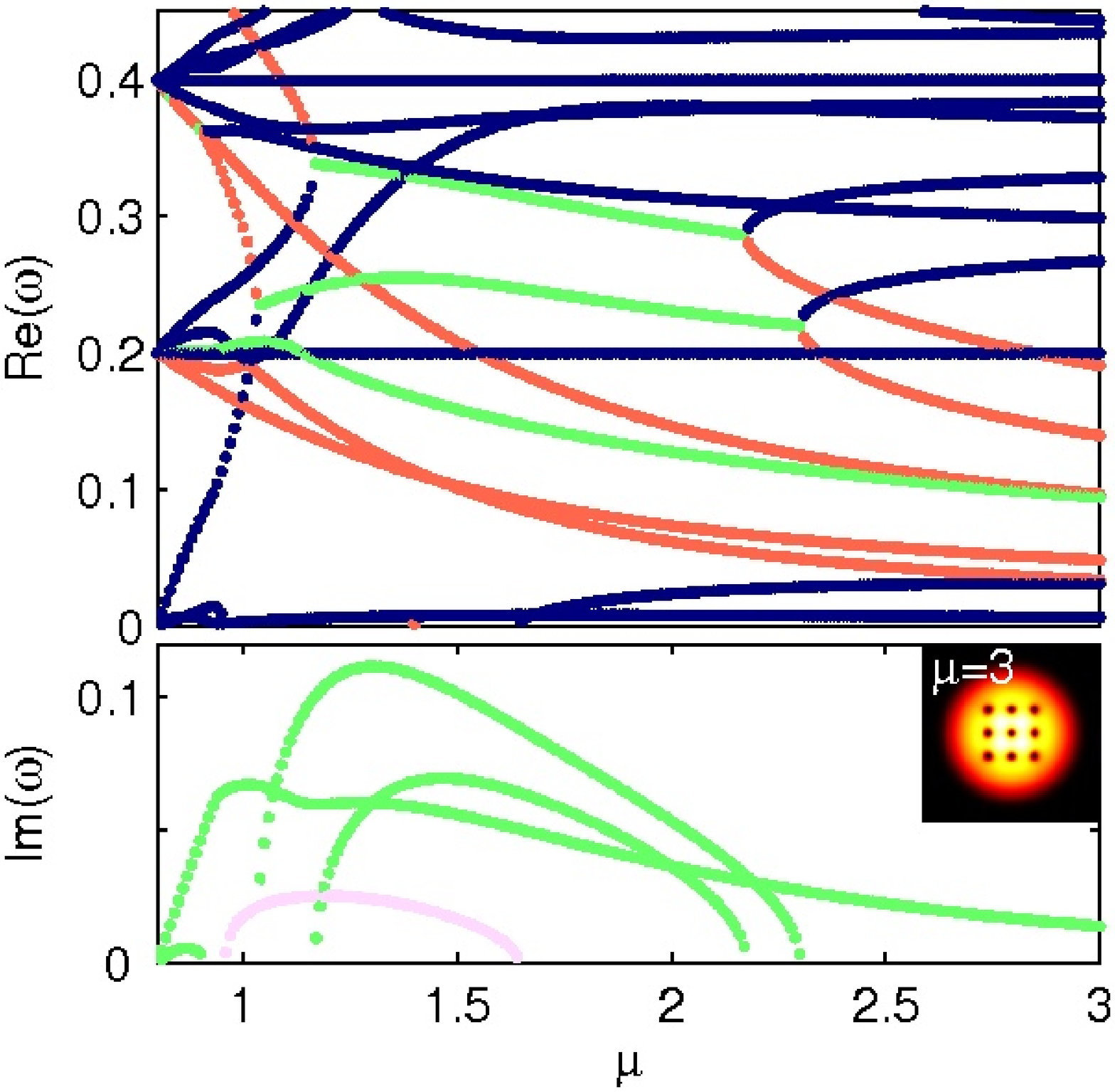} &
\includegraphics[width=4.4cm,height=4.0cm,angle=0]{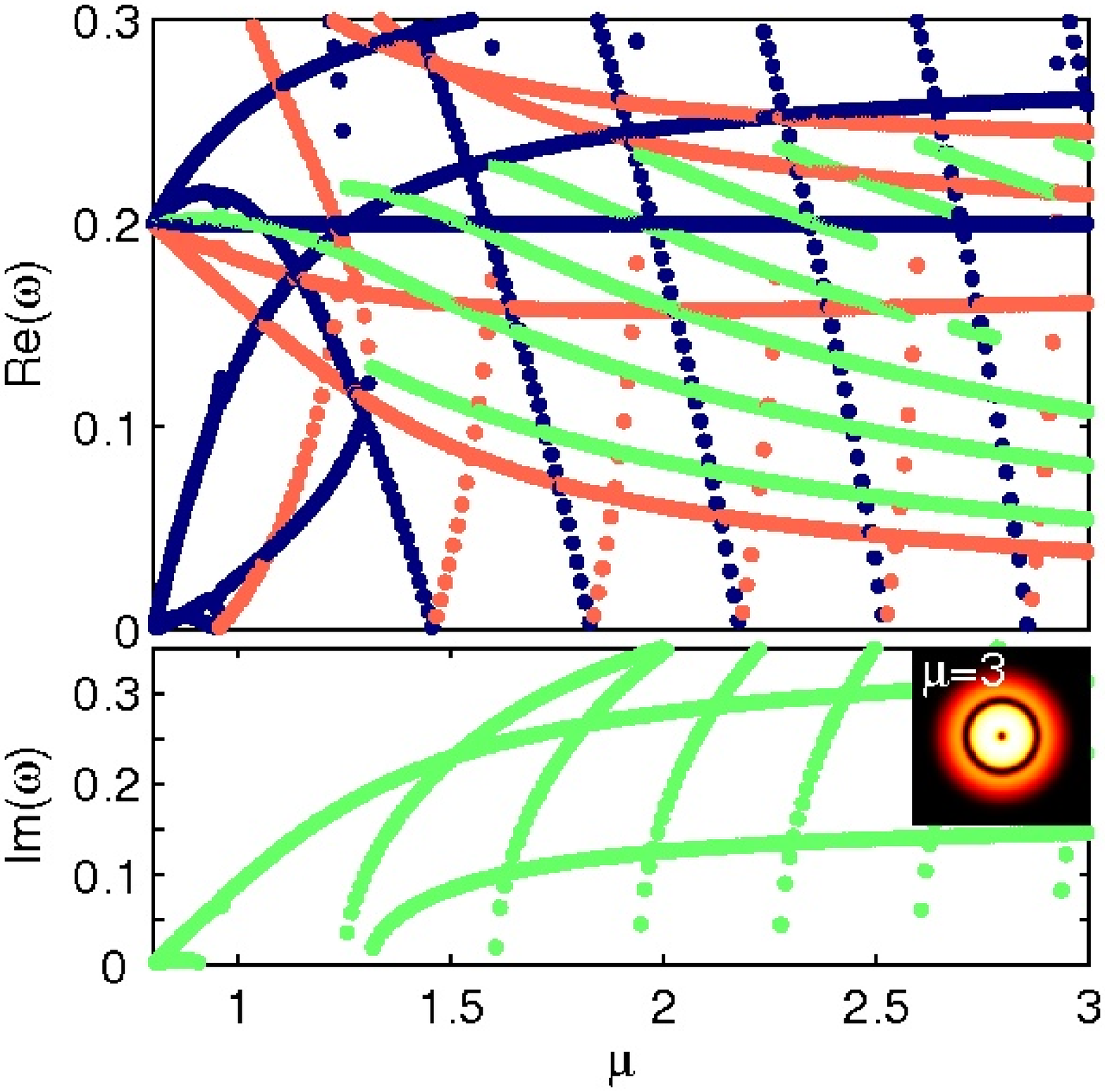} \\[-0.1cm]
\includegraphics[width=4.4cm,height=4.0cm,angle=0]{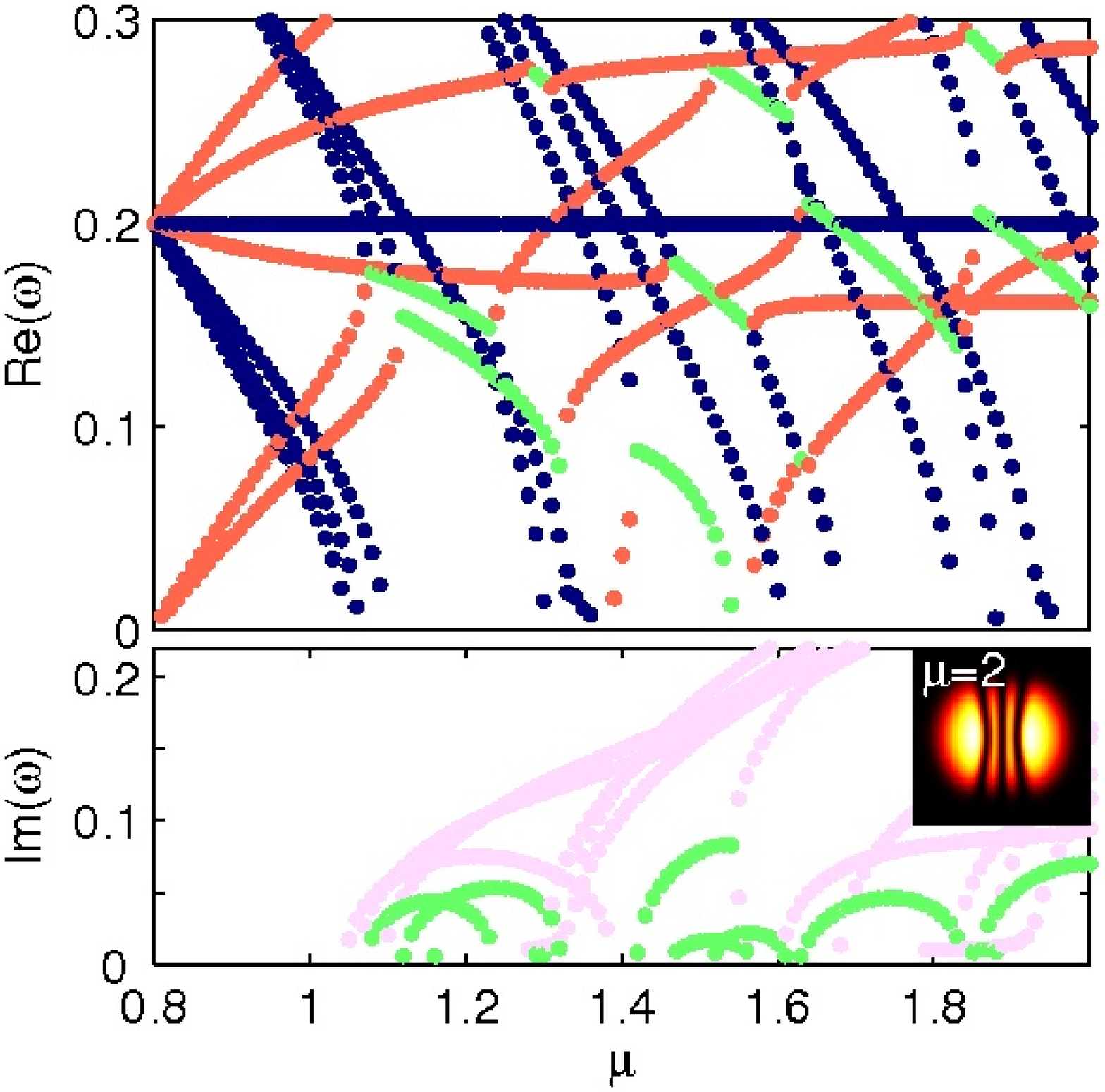} &
\includegraphics[width=4.4cm,height=4.0cm,angle=0]{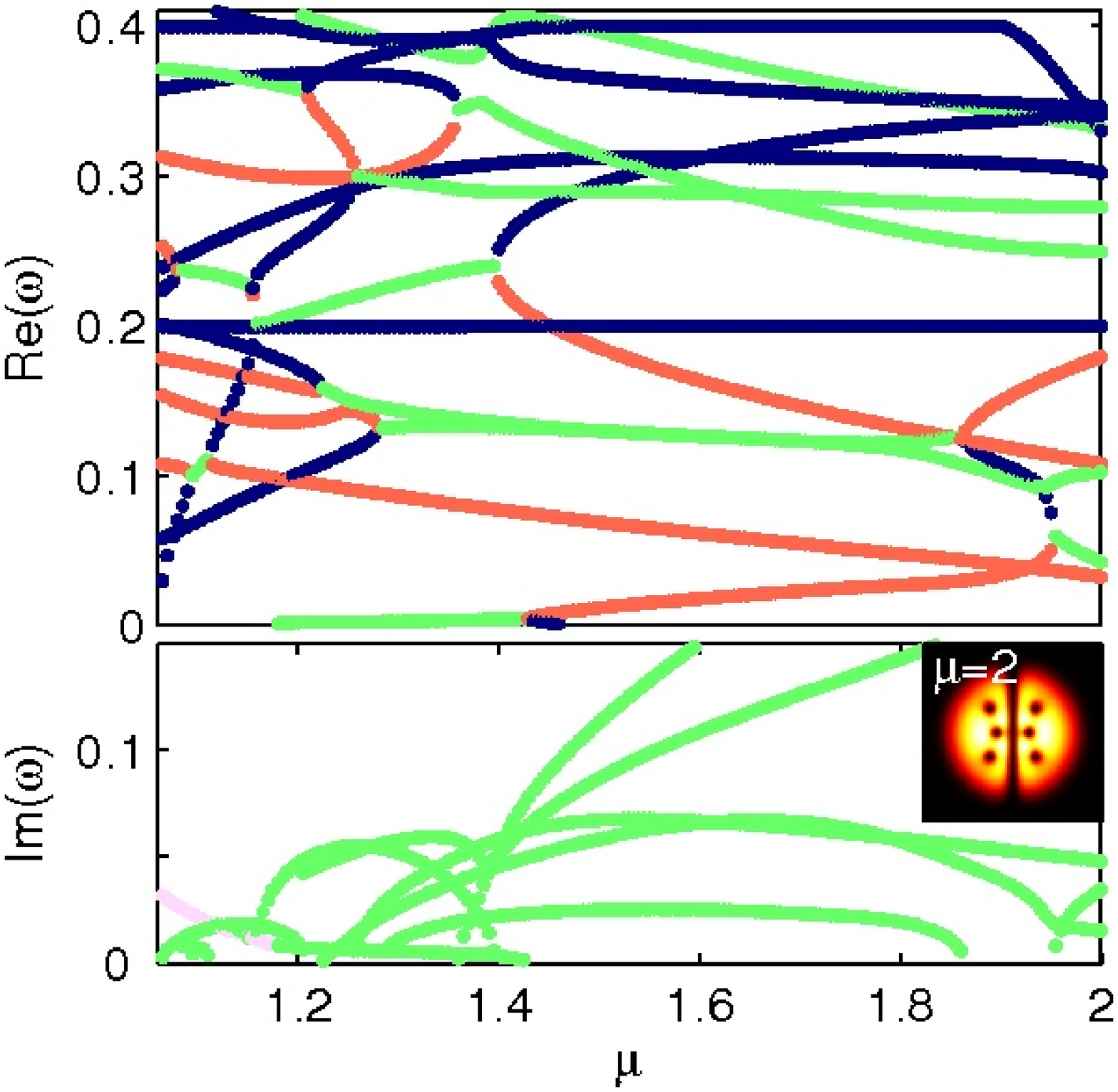}
\end{tabular}
\vspace{-0.5cm}
\caption{(Color online) 
BdG spectra for the bifurcating states depicted in
Fig.~\ref{fig_other}. From left to right and top to bottom:
the nine-vortex state, the ring dark soliton plus vortex, 
the three-dark-soliton stripe (which is stable for low values
of the chemical potential), and 
the mixed soliton-vortex state.
For an explanation of the color codes see Fig.~\ref{fig_BdG_2stripe_cross}.
}
\label{fig_BdG_other}
\end{figure}

In Figs.~\ref{fig_2stripe_cross} and \ref{fig_ring}, 
we explore numerically the above symmetry-breaking events  
for the second excited states and show besides the bifurcation diagram also typical  
profiles of the waveforms, as they result from varying the chemical potential. 
Specifically, Fig.~\ref{fig_2stripe_cross} depicts the bifurcations and
profiles for the states arising from the two dark soliton stripes
(middle column of panels) and
the dark soliton cross (right column of panels).
From the two-stripe soliton bifurcates a six-vortex (6v2) ($2 \times 3$)
state and an eight-vortex (8v2) state ($2 \times 4$) in the considered regime.
From the X-shaped dark soliton cross bifurcates a diagonal six- (6x) and 
eight-vortex (8x) configurations, as well as a state with a vortex 
quadrupole surrounding a central doubly-charged vortex (5x).
We also mention in passing that interesting additional bifurcation events
(collisions and disappearances into ``blue sky'' bifurcations) arise, e.g., 
between the six-vortex cluster deriving from the X-shaped
dark soliton cross and that deriving from the two-stripe soliton
(see blue circle denoted by {\tt A} in Fig.~\ref{fig_2stripe_cross}).
On the other hand,
in Fig.~\ref{fig_ring} we depict the states that bifurcate from the
vortex ring: vortex hexagons (6r) and octagons (8r). 
In this figure we also depict the
vortex quadrupole state (bottom right panel) which 
does not bifurcate from the soliton ring.
In all the cases, good quantitative agreement is found on the critical points 
predicted by the theory and those observed numerically. 

Figures~\ref{fig_BdG_2stripe_cross} and \ref{fig_BdG_ring} complement the 
above 
existence picture with a systematic linear stability analysis of each of the 
corresponding states for Figs.~\ref{fig_2stripe_cross} 
and \ref{fig_ring} respectively.
The two-stripe soliton is dynamically stable near the 
linear limit, a stability inherited by its first symmetry-breaking 
offspring, the six-vortex ($2 \times 3$) state;
while the aligned 
eight-vortex case ($2 \times 4$) is unstable from its existence 
onset (see left column of panels in Fig.~\ref{fig_BdG_2stripe_cross}).
Similarly, the X-shaped dark soliton cross state also
bears imaginary eigenfrequencies for all values of $\mu$ and hence 
its derivative states inherit its dynamical instability
(see right column of panels in Fig.~\ref{fig_BdG_2stripe_cross}). 
On the other hand,
the ring dark soliton is, as was also found earlier \cite{todd,ring1,fr4,herring}, 
always unstable, hence the polygonal vortices that derive from it inherit 
this instability (see Fig.~\ref{fig_BdG_ring}). However, it should be
noted that the instability of such states weakens as the chemical
potential $\mu$ increases.
Note that,
the vortex quadrupole is stable in the linear limit. Since no state bifurcates 
from the vortex quadrupole this stability is generally
preserved (apart from
a small window of oscillatory 
instabilities). 

A general comment on the observed BdG spectra is due here.
We can see that the spectra bear a large number of positive Krein sign
modes (see blue (black) points is all BdG spectra) which appear 
to be asymptoting to appropriate corresponding values in the
large chemical potential limit. In addition, there is a number of
negative Krein sign eigenfrequencies (see orange (dark gray)
points is all BdG spectra) which may lead to 
oscillatory (upon collision with positive sign ones) or exponential
instabilities. Generally, we can comment that the positive 
Krein sign eigenmodes correspond to the ground state ``background''
on which a particular solitonic or vortex (or mixed) state
exists. Furthermore, these eigenfrequencies have a well-defined limit 
when $\mu$ is large as discussed e.g.~in Ref.~\cite{kevrpel} (see
also references therein). On the other hand, the negative Krein
sign modes reflect the excited state nature of the considered
soliton or vortex (or mixed)
states and are the ones which bear the potential for dynamical 
instabilities.
While for solitonic states, we do not have a precise count
on the number of the latter eigenmodes, in the case of multi-vortex
clusters consisting of individual vortices an upper bound
on the maximal order of such (negative Krein) eigenstates
can be given by the number of vortices in the configuration.
This, in turn, also gives an upper bound on the number of 
potentially observable unstable eigenmodes.

Importantly, the present approach can be generalized to higher excited states. 
As merely a small
sample of further exotic configurations that can emerge (some of which can even 
be structurally robust), we mention a 9-vortex cluster (a square grid 
of $3 \times 3$ vortex ``particles''), which emerges from the linear limit 
as $u_{30} + i u_{03}$ and can, therefore, be regarded as a higher 
excited analog to the single vortex and 
the vortex quadrupole states and is expected to be stable in
the vicinity of that limit 
(at least with respect to purely imaginary [i.e., non-oscillatory] 
instabilities, see below).
On the other hand, there are bifurcations from that
state including the bifurcation of a ring with a vortex in its center (which was again discussed 
e.g., in Ref.~\cite{herring}).
Another state that exists and should be robust near the 
linear limit is a three-soliton-stripe. In fact, from such a state 
bifurcations again emerge due to the mixture with states such 
as $\phi_1=u_{0m}$ with $m \geq 4$, but also with $\phi_{1}=u_{1m}$
with $m \geq 3$. We focus briefly on the latter, which is theoretically
predicted to occur at $\mu_{\rm cr}^{\rm 6vs}=1250 \Omega/239$
(see vertical dashed line in Fig.~\ref{fig_other}), because it gives rise
to yet another novel type of state, namely a mixed state between  
vortices and a dark soliton: this soliton-vortex mixed state has a dark 
soliton 
stripe nodal line plus two additional lines over each of which three vortices
reside. Similar states with 8-, 10- etc. vortices beside the soliton 
also exist, arising through subsequent bifurcations.

To corroborate the above theoretical predictions we show in 
Figs.~\ref{fig_other} and \ref{fig_BdG_other} (again for $\Omega=0.2$) 
some prototypical examples of states that 
emerge from the third excited linear branches
and their corresponding BdG stability spectra.
The 9-vortex ``crystal'' is stable close to the linear limit in the sense that 
its BdG spectrum possesses no imaginary mode. 
However there is a complex mode inducing an oscillatory instability. 
At $\mu\approx0.95$ the vortex plus ring soliton state bifurcates from 
the 9-vortex ``crystal'' inducing a complex mode. 
However, the latter is small and vanishes again at $\mu\approx1.6$. 
The BdG spectrum of the vortex plus ring soliton state contains no purely
imaginary mode but many complex modes. Note that the eigenvalue spectrum 
looks fairly similar to the ring dark soliton case, but the imaginary part 
of the modes is not due to imaginary modes which are created by bifurcations. 
In this case, modes with positive energy cross zero and thus obtain a 
negative energy. These negative energy modes then collide with modes 
with positive energy and create the complex modes.
Moreover, we show the three-soliton-stripe, as well as the mixed 
soliton-six-vortex cluster state emerging from its first
bifurcation event in Fig.~\ref{fig_other}. The three-soliton state 
is stable near the linear limit
(of $\mu=0.8$) but becomes destabilized at $\mu=1.05$ (and 
then further so at $1.06$, $1.1$ and $1.3$). The first bifurcation 
gives rise to the very weakly unstable soliton-vortex state 
predicted above; the critical point for this bifurcation 
is found to be at $\mu=1.034$, once again in very good
agreement with the full numerical result. 
The present approach can naturally be extended to a multitude of additional 
states which, however, are expected to be dynamically unstable; therefore, 
having presented the most fundamental ones, we will not proceed further 
with such considerations.

\section{Conclusions and Future Directions}
\label{Sec4}

In conclusion, in the present work we have shown 
that a detailed understanding of emergent vortex 
cluster states is possible through 
a near-linear-limit approach. This involves identifying the possible linear
states and tracking systematically the symmetry-breaking bifurcations
that can arise from them. This allowed us not only to discuss a
cascade of bifurcations from the first excited state (in the
form of aligned vortex clusters), but to also reveal a broad
class of states emerging from the second- and third-excited
states. These were not only rectilinear states (with ``soliton-type''
stripes), but also soliton rings and rings with vortices, 
vortex quadrupoles or nine-vortex-crystals, as well as various 
clusters of vortices that derive from some of these states, 
including vortex hexagons, octagons, decagons, $n \times m$ 
states (of $m$ vortices sitting along $n$ stripes), 
soliton-vortex states, and so on.

We were also able, based on the general bifurcation structure
of the problem, to reveal which ones among these states are
expected to be most robust, such as the vortex dipole or quadrupole,
and some of the emerging vortex clusters, such as the $2 \times 3$
or the soliton-vortex state.

It would be especially interesting to extend this picture to different
dimensions. On the one hand, one can consider effects of
anisotropy (by changing the strengths of the two 
different trapping directions).
This should enable a ``dimension-transcending'' picture, 
as extreme anisotropies should allow to observe how the system 
transitions between quasi-one- and genuinely two-dimensional dynamics. 
On the other hand, it would be particularly relevant and interesting 
to attempt to extend such considerations into three-dimensional 
settings, and understand how relevant ideas generalize
and potentially give rise to structures such as vortex rings that 
have been observed in pertinent experiments \cite{ander,hau}.
Such investigations are currently in progress. 

\section*{Acknowledgments}
%
P.G.K. gratefully acknowledges support from NSF-DMS-0349023 (CAREER), 
NSF-DMS-0806762 and the Alexander von Humboldt Foundation.
The work of D.J.F. was partially supported by the
Special Account for Research Grants of the University of Athens.
R.C.G. acknowledges support from  NSF-DMS-0806762.
The authors also gratefully acknowledge David Hall for 
numerous enlightening discussions on experimental aspects of this theme.


\end{document}